\documentclass[prb,10pt]{revtex4-1}

\usepackage{lineno,hyperref}
\usepackage{bm}
\usepackage{mathtools}
\usepackage{amsmath,bm}
\usepackage{amssymb}
\usepackage{subfigure}
\usepackage{setspace}
\usepackage{dcolumn}
\usepackage{color}
\usepackage{mathptmx}
\usepackage{relsize}
\usepackage{color}

\newcommand{\sqroot}[1]{{\mathlarger{\sqrt{\mathsmaller{\mathsmaller{#1}}}}}}

\modulolinenumbers[5]

\begin{document}

\title{Exceptional points and enhanced sensitivity in PT-symmetric continuous elastic media }

\author{Matheus I. N. Rosa}
\author{Matteo Mazzotti}
\author{Massimo Ruzzene}
\affiliation{Department of Mechanical Engineering, University of Colorado Boulder, USA}

%
%

\date{\today}

\begin{abstract}
We investigate non-Hermitian degeneracies, also known as exceptional points, in continous elastic media, and their potential application to the detection of mass and stiffness perturbations. Degenerate states are induced by enforcing parity-time symmetry through tailored balanced gain and loss, introduced in the form of complex stiffnesses and may be implemented through piezoelectric transducers. Breaking of this symmetry caused by external perturbations leads to a splitting of the eigenvalues, which is explored as a sentitive approach to detection of such perturbations. Numerical simulations on one-dimensional waveguides illustrate the presence of several exceptional points in their vibrational spectrum, and conceptually demonstrate their sensitivity to point mass inclusions. Second order exceptional points are shown to exhibit a frequency shift in the spectrum with a square root dependence on the perturbed mass, which is confirmed by a perturbation approach and by frequency response predictions. Elastic domains supporting guided waves are then investigated, where exceptional points are formed by the hybridization of Lamb wave modes. After illustrating a similar sensitivity to point mass inclusions, we also show how these concepts can be applied to surface wave modes for sensing crack-type defects. The presented results describe fundamental vibrational properties of PT-symmetric elastic media supporting exceptional points, whose sensitivity to perturbations goes beyond the linear dependency commonly encountered in Hermitian systems. The findings are thus promising for applications involving sensing of perturbations such as added masses, stiffness discontinuities and surface cracks. 
\end{abstract}

\maketitle

%
%

\section{Introduction}

In the broad context of wave physics and related areas, a recent focus of the scientific community at large has been the exploration of non-Hermitian (NH) systems~\cite{ashida2020non}. Broadly speaking, NH systems are non-conservative due to interactions with the environment producing gain and/or loss. As such, the eigenfrequencies of NH systems are generally complex, precluding many of the well-known properties of Hermitian (conservative) systems to be directly applied. For example, the topological properties of NH systems are remarkably different from that of their Hermitian counterparts~\cite{lee2016anomalous,xiong2018does}, motivating re-classifications of topological phases for NH systems to be recently proposed~\cite{gong2018topological,shen2018topological,kawabata2019symmetry}. Recently, these concepts have been exploited to produce one-way wave amplification and edge localization of bulk modes (called the NH skin effect) in mechanical metamaterials with feedback interactions~\cite{ghatak2019observation,brandenbourger2019non,rosa2020dynamics}. In this context, recent studies focus on a particular class of NH systems which preserve Parity-Time (PT) symmetry and exhibit exceptional points~\cite{longhi2018parity,el2018non,miri2019exceptional}. The interest in these systems has early roots in the seminal work by Bender and Boettcher~\cite{bender1998real}, where it was shown that PT-symmetric NH Hamiltonians may exhibit purely real spectra. Exceptional points are special degeneracies where two or more eigenfrequencies and eigenvectors system coalesce, defining a transition from a phase where eigenfrequencies are purely real, to one where they are complex conjugates~\cite{bender2013observation}. Among the intriguing properties of exceptional points one finds unconventional phenomena such as unidirectional invisibility~\cite{lin2011unidirectional,fleury2015invisible}, single-mode lasers~\cite{feng2014single}, and enhanced sensitivity to perturbations~\cite{hodaei2017enhanced,chen2017exceptional,xiao2019enhanced}. Such sensitivity is investigated here for continuous elastic media. This work contributes to recent efforts in exploring the role of PT symmetry in acoustics and mechanics~\cite{zhu2014p,christensen2016parity,liu2018unidirectional}. Notable applications include the observation of the asymmetric scattering properties around exceptional points in 1D waveguides~\cite{fleury2015invisible,wu2019asymmetric,hou2018tunable}, and second-order topological insulators demonstrated in acoustics~\cite{zhang2019non,
lopez2019multiple}. More recently, exceptional points were experimentally demonstrated as vibrating modes of an elastic plate hosting two mechanical oscillators with tailored losses~\cite{dominguez2020environmentally}, while an optomechanical accelerometer based on the enhanced sensitivity around exceptional points was also proposed~\cite{kononchuk2020orientation}. Also, based on parallels between elastodynamics and NH quantum mechanics, exceptional points and their sensitivity to point masses have been recently illustrated for bi-material elastic slabs~\cite{shmuel2020linking}. These works illustrate the potential of elastic domains for hosting exceptional points with enhanced sensing capabilities. However, a treatment from a fundamental structures perspective is still missing, along with potential implementations related to sensing and detection. Towards bridging this gap, we investigate 1D and 2D PT symmetric elastic domains and illustrate the formation of exceptional points arising from the introduction of balanced gain and loss elements. In particular, we first investigate 1D elastic waveguides featuring a PT symmetric pair of ground springs with complex stiffnesses, and illustrate how a large number of exceptional points naturally appear in their vibrational spectra. We investigate their sensitivity to perturbations in the form of point mass inclusions, and demonstrate both numerically and through a perturbation approach that the sensitivity has a leading term of square root order, as expected of second-order exceptional points~\cite{chen2017exceptional,shmuel2020linking}. These finding are useful when extended to 2D elastic domains, where gain and loss may be implemented through piezoelectric transducers to detect mass inclusions and surface cracks. In this context, our results open new avenues in the area of dynamic-based non-destructive testing. Changes in modal properties have long been investigated as tools to detect structural changes resulting for example from the onset of cracks. Methods based on shifts in natural frequencies provide in principle a convenient detection scheme that requires limited sensing, but they have broadly shown strong sensitivity limitations. Other techniques have explored monitoring of mode shapes and curvature shapes~\cite{sharma2006perturbation}, which while promising, suffer from high spatial measurement resolution requirements. Examples that exploit the perturbation of modal parameters, such as those caused by cracks in elastic beams and plates are illustrated in~\cite{luo1997integral,sharma2006perturbation}. A review of other well-establish methods for sensing and damage detection can be found in~\cite{staszewski2004health,giurgiutiu2007structural}. Also, guided waves have been employed for damage detection in plates~\cite{ruzzene2007frequency} and composite materials~\cite{kessler2002damage}, while surface acoustic waves have been employed for sensing liquid viscosity~\cite{jakoby1998viscosity} and surface mass loading~\cite{du1996study}. Several of these contributions exploit a linear sensitivity with respect to a perturbation parameter, which may potentially be improved in the context of exceptional points with enhanced sensitivity of square root leading order. Therefore, the findings presented in this paper may open new avenues for the general exploration of PT symmetry, exceptional points and their sensitivity in continuous elastic media, with potential applications ranging from sensors to novel structural health monitoring strategies. 

This paper is organized as follows: following this introduction, section~\ref{Rodssec} presents the analysis of PT-symmetric 1D waveguides, describing the formation of exceptional points and their sensitivity using a perturbation approach. Section~\ref{2Dsec} then presents the analysis of 2D elastic domains with PT-symmetric pairs of piezoelectric patches, where exceptional points and their sensitivity to perturbations is numerically investigated. Finally, section~\ref{Conclusionsec} summarizes the main findings of the work and briefly outlines future research directions.

\section{PT symmetric elastic rods}\label{Rodssec}

\begin{figure}[t!]
\includegraphics[width=0.9\textwidth]{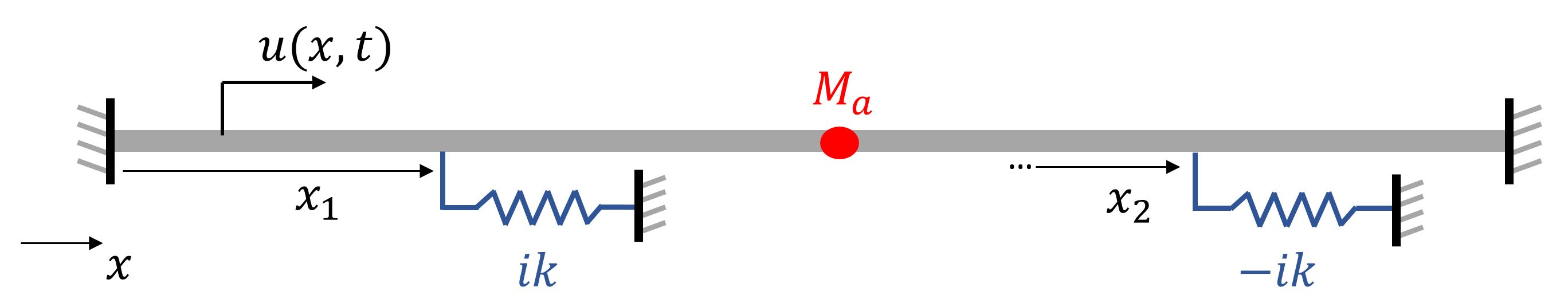}
\centering
\caption{Schematic of elastic rod with PT Symmetric pair of ground springs and a point mass $M_a$ attached to the center.}
\label{Figrodschematic}
\end{figure}

We begin our study by considering a 1D elastic waveguide equipped with a pair of ground springs (Fig.~\ref{Figrodschematic}). We employ conceptual complex spring constants, whose imaginary components with opposite signs induce gain and loss in equal proportions, thus making the structure PT symmetric~\cite{dominguez2020environmentally}. We illustrate how exceptional points emerge in such a system, and investigate their sensitivity to a perturbation in the form of a point mass inclusion ($M_a$). 

\subsection{Governing equations and approximate solution approach}
The equation governing the motion of the waveguide can be generally expressed as~\cite{meirovitch1975elements}:
\begin{equation}\label{EqGov1}
\mathcal{L}[u(x,t)]-m(x)\dfrac{\partial^2u(x,t)}{\partial t^2}=q(x,t),
\end{equation}
where $u(x,t)$ is the longitudinal displacement, $q(x,t)$ is the external loading and $m(x) = \rho A + M_a \delta(x-x_m)$ is the linear mass. Here, $\rho$ and $A$ respectively denote the mass density and cross-sectional area, while $M_a$ is the mass at location $x=x_m$ defined by the $\delta$ function. Assuming longitudinal motion, the linear differential operator $\mathcal{L}$ is given by:
\begin{equation}\label{eqoperator}
\mathcal{L}=EA\dfrac{\partial^2}{\partial x^2} - \sum_{s=1}^S k_s \delta(x-x_s),
\end{equation}
where $E$ is the Young's modulus, $k_s$ and $x_s$ are respectively the spring constant and the location of the $s$-th ($s=1,..,S$) ground spring. While we here focus our attention to axial motion, this formulation lends itself to the analysis of other wave motion, such as flexural (transverse) vibrations for which the operator $\mathcal{L}$ includes a fourth-order derivative~\cite{Pal_2019}. Numerical results for flexural vibrations of elastic beams are presented at the end of this section.

We seek for approximate solutions to Eqn.~\eqref{EqGov1} by expressing the axial displacement through a set of $N$ comparison functions $\phi_n(x)$
\begin{equation}\label{equexpand}
u(x)=\sum_{n=1}^N u_n \phi_n(x), \,\,\, \phi_n(x)= \sin \left( \frac{\pi n x}{L} \right), \qquad n=1, ..., N,
\end{equation}
for a rod of length $L$, that is fixed at bound ends, \emph{i.e.} $u(0)=u(L)=0$.

Assuming an external harmonic load $q(x,t)=q(x)e^{i\omega t}$, the application of Galerkin's method~\cite{meirovitch1975elements} leads to a set of $N$ algebraic equations, which can be expressed in the following matrix form:
\begin{equation}\label{eqforced}
(\mathbf{K}-\omega^2\mathbf{M})\mathbf{u}=\bm{q}
\end{equation}
where $\mathbf{u}=[u_1, u_2, ..., u_N]^T$, while $\mathbf{K}$ and $\mathbf{M}$ are the $N \times N$ mass and stiffness matrices, whose $i,j$-th entries are given by:
\begin{align}
k_{ij}&= \int_0^L \mathcal{L}[\phi_i(x)]\phi_j(x) dx = EA\left( \frac{i^2\pi^2}{2L} \right)\delta_{ij}+\sum_{s=1}^Sk_s\phi_i(x_s)\phi_j(x_s) \nonumber \\
m_{ij}&=\int_0^Lm(x)\phi_i(x)\phi_j(x)dx = \frac{\rho AL}{2} \delta_{ij} + M_a\phi_i(x_m)\phi_j(x_m),
\end{align}
with $\delta_{ij}$ denoting the Kronecker delta. Also in Eqn.~\eqref{eqforced}, $\bm{q}=[q_1, q_2, ..., q_N]^T$ is the external load projected in the basis of the comparison functions, i.e. $q_n=\int_0^Lq(x)\phi_ndx$. For the purposes of the present work, we consider a point force $f$ applied at a location $x=x_f$, i.e. $q(x,t)=f(t) \delta(x-x_f)$, which gives $q_n(t)=f(t)\phi_n(x_f)$. 

%

\subsection{Perturbation approach}
We carry out a perturbation approach to predict the frequency splitting at an exceptional point due to the perturbation associated with the point mass inclusion $M_a$. While we consider perturbations in the mass matrix $\mathbf{M}$ only, similar derivations can be carried out for stiffness perturbations that affect matrix $\mathbf{K}$. We consider a first order expansion of the mass matrix around $\epsilon=0$, where $\epsilon=M_a/(\rho A L)$ is the non-dimensional parameter associated with the added mass. As such, the eigenvalue problem associated with the homogeneous form of Eqn.~\eqref{eqforced} is rewritten as
\begin{equation}\label{EigPert}
\mathbf{K}\mathbf{u}=\omega^2(\mathbf{M}_0+\epsilon\mathbf{M}_1)\mathbf{u},
\end{equation}
where $\mathbf{M}_0=\mathbf{M}|_{\epsilon=0}$, while $\mathbf{M}_1=\partial\mathbf{M}/\partial\epsilon|_{\epsilon=0}$ is the contribution due to the added mass, i.e. $m_{1_{ij}}=(\rho A L) \phi_i(x_m)\phi_j(x_m)$. 

An exceptional point is a degenerate eigenvalue of algebraic multiplicity $r$ and geometric multiplicity $1$, producing a single linearly independent eigenvector. The perturbation around such type of degeneracy follows a Newton-Puiseux series of leading order $\epsilon^{1/r}$~\cite{seyranian2003multiparameter}, in contrast with common degeneracies producing two linearly dependent eigenvectors where the perturbation follows with leading order $\epsilon$, \emph{i.e.} $\mathcal{O}(\epsilon)$. We here focus on second order exceptional points with $r=2$, resulting in the following expansions of eigenvectors and eigenfrequencies
\begin{align}\label{PerturbationExpansions}
\mathbf{u}& = \mathbf{u}_0 + \mathbf{u}_1\epsilon^{1/2} + \mathbf{u}_2\epsilon^{1} + \mathbf{u}_3\epsilon^{3/2} + \mathbf{u}_4\epsilon^{2} + ... \\
\bm{\omega}& = \bm{\omega}_0 + \bm{\omega}_1\epsilon^{1/2} + \bm{\omega}_2\epsilon^{1} + \bm{\omega}_3\epsilon^{3/2} + \bm{\omega}_4\epsilon^{2} + ...
\end{align}

Substitution of the series expansions into Eqn.~\eqref{EigPert} yields the following set of ordered equations
\begin{align}
&\epsilon^0 : \hspace{3.5mm}  \mathbf{K}\mathbf{u}_0 = \omega_0^2\mathbf{M}_0\mathbf{u}_0 \\
&\epsilon^{1/2} : \hspace{1mm} (\mathbf{K} - \omega_0^2\mathbf{M}_0)\mathbf{u}_1= 2\omega_0\omega_1\mathbf{M}_0\mathbf{u}_0 \\
&\epsilon^1 : \hspace{3.5mm} (\mathbf{K}-\omega_0^2\mathbf{M}_0)\mathbf{u}_2 = \omega_0^2\mathbf{M}_1\mathbf{u}_0+2\omega_0\omega_1\mathbf{M}_0\mathbf{u}_1 + (\omega_1^2+2\omega_0\omega_2)\mathbf{M}_0\mathbf{u}_0 \\
&\epsilon^{3/2} : \hspace{1mm} (\mathbf{K}-\omega_0^2\mathbf{M}_0)\mathbf{u}_3 = 2\omega_0\omega_1\mathbf{M}_0\mathbf{u}_2+2\omega_0\omega_1\mathbf{M}_1\mathbf{u}_0+(\omega_1^2+2\omega_0\omega_2)\mathbf{M}_0\mathbf{u}_1 \\
& \hspace{75mm} +\omega_0^2\mathbf{M}_1\mathbf{u}_1+(2\omega_0\omega_3+2\omega_1\omega_2)\mathbf{M}_0\mathbf{u}_0 \nonumber 
\end{align}

Note that to obtain up to the second correction $\omega_2$, equations up to $\epsilon^{3/2}$ are needed. The $\epsilon^0$ equation corresponds to the eigenvalue problem of the unperturbed system, i.e. the rod with no added mass. This eigenvalue problem is solved numerically as a function of the spring constant $k$ to find the existence of exceptional points. A second order exceptional point produces a double eigenvalue $\omega_0$ and a single eigenvector $\mathbf{u}_0$. The corresponding left-eigenvector $\mathbf{v}_0$ satisfying $\mathbf{v}_0^{H}\mathbf{K}=\omega_0^2\mathbf{v}_0^{H}\mathbf{M}_0$ is also obtained numerically, where $( )^H$ denotes the conjugate transpose. It is useful to consider the Jordan chain of length $2$ associated with the degeneracy~\cite{seyranian2003multiparameter}, i.e. $\{ \mathbf{u}_0, \mathbf{w}_0 \}$ and $\{ \bm{v_0}, \mathbf{z}_0 \}$, where $\mathbf{w}_0$ and $\mathbf{z}_0$ are associated (or generalized rank 2) right and left eigenvectors satisfying the equations
\begin{equation}\label{associatedeig}
\mathbf{K}\mathbf{w}_0=\omega_0^2\mathbf{M}_0\mathbf{w}_0+\mathbf{M}_0\mathbf{u}_0, \qquad \qquad \mathbf{z}_0^{H}\mathbf{K}=\omega_0^2\mathbf{z}_0^{H}\mathbf{M}_0 + \mathbf{v}_0^H\mathbf{M}_0,
\end{equation}
with the following normalization conditions
\begin{equation}\label{normalization}
\mathbf{v}_0^H\mathbf{M}_0\mathbf{w}_0=\mathbf{z}_0^H\mathbf{M}_0\mathbf{u}_0=1, \qquad \qquad \mathbf{v}_0^H\mathbf{M}_0\mathbf{u}_0=\mathbf{z}_0^H\mathbf{M}_0\mathbf{w}_0=0.
\end{equation}

A choice of associated eigenvectors $\{ \mathbf{w}_0, \mathbf{z}_0 \}$ is determined numerically by using the pseudo-inverse of the matrix $\mathbf{K}-\omega_0^2\mathbf{M}_0$. While the Jordan chain is not unique~\cite{seyranian2003multiparameter}, for a given choice of right eigenvectors $\{ \mathbf{u}_0, \mathbf{w}_0 \}$ the left eigenvectors $\{ \mathbf{v}_0, \mathbf{z}_0 \}$ are uniquely determined by the normalization conditions stated in Eqn.~\eqref{normalization}. To determine the perturbed eigenvector $\mathbf{u}$ uniquely, it is convenient to consider the normalization $\mathbf{z}_0^{H}\mathbf{M}_0\mathbf{u}=1$. Since $\mathbf{z}_0^{H}\mathbf{M}_0\mathbf{u}_0=1$ from Eqn.~\eqref{normalization}, this gives the following conditions for the eigenvector perturbations:
\begin{equation}\label{normalization2}
\mathbf{z}_0^{H}\mathbf{M}_0\mathbf{u}_i = 0 \qquad \mbox{,} \quad i=1, 2, ...
\end{equation}

Starting with the $\epsilon^{1/2}$ equation, we remark that the matrix operator $\mathbf{K}-\omega_0^2\mathbf{M}_0$ is singular with rank $N-1$ due to the degeneracy at $\omega_0$. To circumvent this issue, the normalization condition for $\mathbf{u}_1$ given by Eqn.~\eqref{normalization2} is pre-multiplied by $\mathbf{v}_0$ and added to the $\epsilon^{1/2}$ equation, yielding
\begin{equation}\label{eps1/2}
\bm{G}\mathbf{u}_1=2\omega_0\omega_1\mathbf{M}_0\mathbf{u}_0,
\end{equation}
where $\bm{G}=\mathbf{K}-\omega_0^2\mathbf{M}_0+\mathbf{v}_0\mathbf{z}_0^H\mathbf{M}_0$ becomes a non-singular matrix due to the addition of the term $\mathbf{v}_0\mathbf{z}_0^H\mathbf{M}_0$~\cite{seyranian2003multiparameter}. The same procedure applied to the $\mathbf{w}_0$ equation (Eqn.~\eqref{associatedeig}) gives $\bm{G}\mathbf{w}_0=\mathbf{M}_0\mathbf{u}_0$, which when compared to Eqn.~\eqref{eps1/2} (and noting that $\bm{G}$ is non-singular) gives
\begin{equation}\label{equ1}
\mathbf{u}_1=2\omega_0\omega_1\mathbf{w}_0.
\end{equation}

Substitution of Eqn.~\eqref{equ1} into the $\epsilon^1$ equation, left multiplication by $\mathbf{v}_0^H$ and considering the normalization conditions described previously yields the $\epsilon^{1/2}$ frequency correction:
\begin{equation}\label{omg1}
\omega_1= \pm \frac{1}{2} \sqroot{-\mathbf{v}_0^H\mathbf{M}_1\mathbf{u}_0}
\end{equation}

Next, we establish the $\epsilon^1$ order correction $\omega_2$. To that end, the $\epsilon^1$ equation is first multiplied by $\mathbf{z}_0^H$ from the left, which upon normalization gives:
\begin{equation}\label{u2eq1}
\mathbf{v}_0^{H}\mathbf{M}_0\mathbf{u}_2=\omega_0^2\mathbf{z}_0^{H}\mathbf{M}_1\mathbf{u}_0 + \omega_1^2 + 2\omega_0\omega_2.
\end{equation}

Now, multiplying the $\epsilon^{3/2}$ equation by $\mathbf{v}_0^{H}$ from the left, and using Eqn.~\eqref{equ1}, Eqn.~\eqref{omg1} and Eqn.~\eqref{u2eq1}, along with normalization conditions, gives
\begin{equation}\label{omg2}
\omega_2 = - \frac{1}{8\omega_0}\mathbf{v}_0^H\mathbf{M}_1\mathbf{u}_0 - \frac{\omega_0}{4}(\mathbf{z}_0^H\mathbf{M}_1\mathbf{u}_0+\mathbf{v}_0^H\mathbf{M}_1\mathbf{w}_0)
\end{equation}

According to the equations above, we conclude that the degeneracy at $\omega_0$ initially splits symmetrically with respect to $\omega_0$ with a $\epsilon^{1/2}$ dependence due to the $\omega_1$ corrections, which are equal and opposite in sign (Eqn.~\eqref{omg1}). For higher $\epsilon$ values a linear correction $\omega_2$ (depending on $\epsilon^1$) becomes relevant, which is equal for both branches (Eqn.~\eqref{omg1}). The difference between the frequency of the two branches considering only the first two perturbation orders is therefore expressed as 
\begin{equation}\label{eqsplit}
\Delta\Omega=\epsilon^{1/2}\sqroot{-\mathbf{v}_0^H\mathbf{M}_1\mathbf{u}_0} \quad,
\end{equation}
since the $\omega_1$ correction is the same for both branches. As further described in the next section, the frequency splitting $\Delta\Omega$ is a parameter commonly used for sensing, which is said to be \textit{enhanced} due to the dependence with $\epsilon^{1/2}$~\cite{hodaei2017enhanced,chen2017exceptional,kononchuk2020orientation,shmuel2020linking}.

\subsection{Numerical results and analysis}

The elastic waveguide has two ground springs attached to locations $x_1$ and $x_2$, with spring constants $k_1=ik$ and $k_2=-ik$ (Fig.~\ref{Figrodschematic}). To guarantee PT symmetry, we consider $x_2=L-x_1$, such that the locations of the ground springs are symmetric with respect to the center, where the mass $M_a$ is attached ($x_m=L/2$). The role of parity is to produce spatial inversions, say with respect to the center $x=L/2$, which when combined with time reversal ($i \to -i$) produces the same structure and operator, confirming the PT-symmetry. As expected~\cite{bender1998real}, we find that the eigenvalues of the PT-symmetric rod are either real or come in complex conjugate pairs, and exceptional points are found defining a transition from the first case to the latter as the spring constant $k$ is varied. In our numerical simulations, we consider $x_1=0.2L$ and a set of $N=400$ basis functions in Eqn.~\eqref{equexpand}. Results are displayed in terms of non-dimensional frequency $\Omega=\omega/\omega_0$, with $\omega_0=\sqrt{E/(\rho L^2)}$, and non-dimensional ground spring stiffness $\gamma=k/k_0$, where $k_0=EA/L$ is a measure of the rod stiffness. 

\begin{figure}[b!]
	\centering
	\subfigure[]{\includegraphics[height=0.275\textwidth]{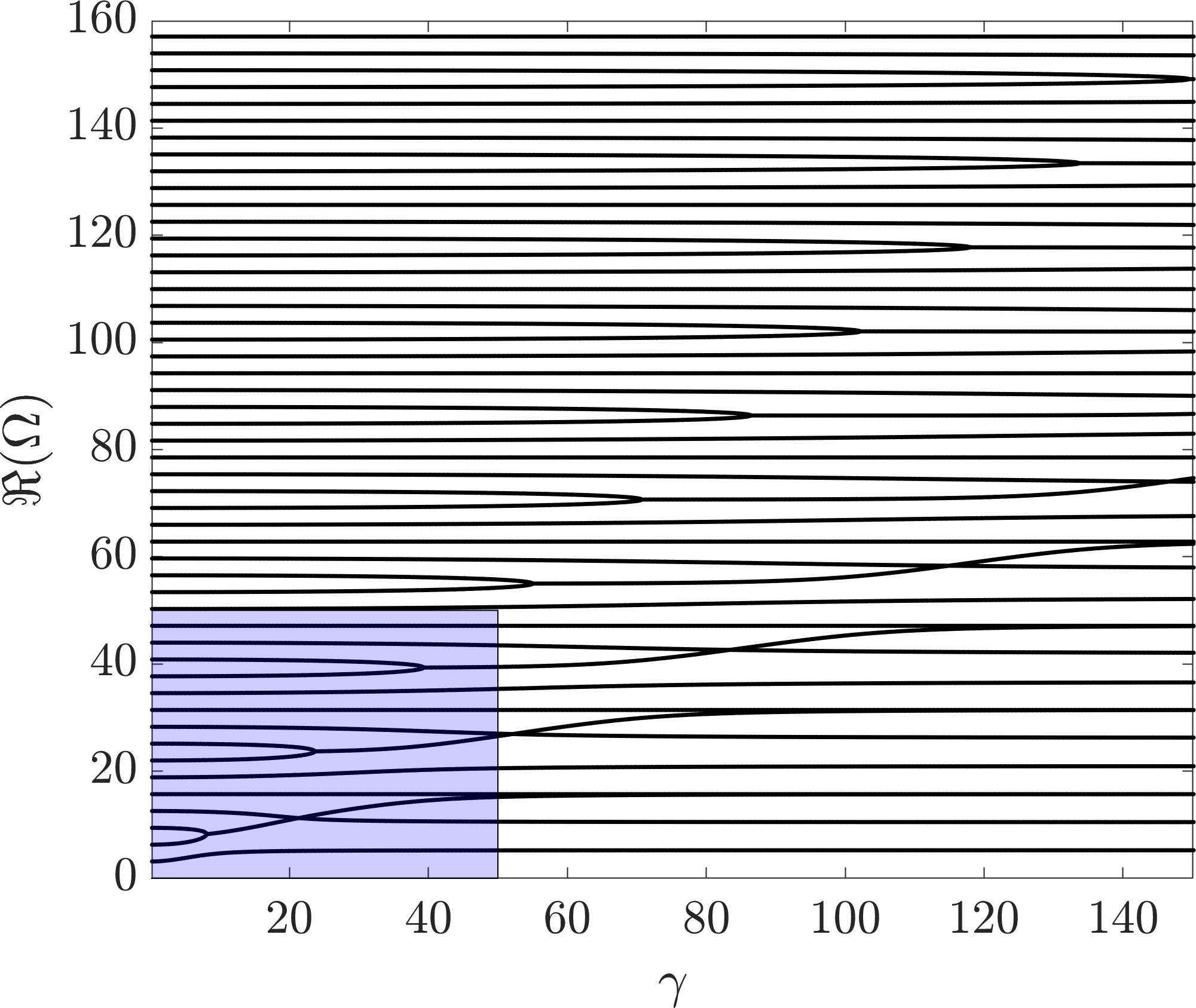}\label{Figrods2a}}\hspace{1mm}
	\subfigure[]{\includegraphics[height=0.275\textwidth]{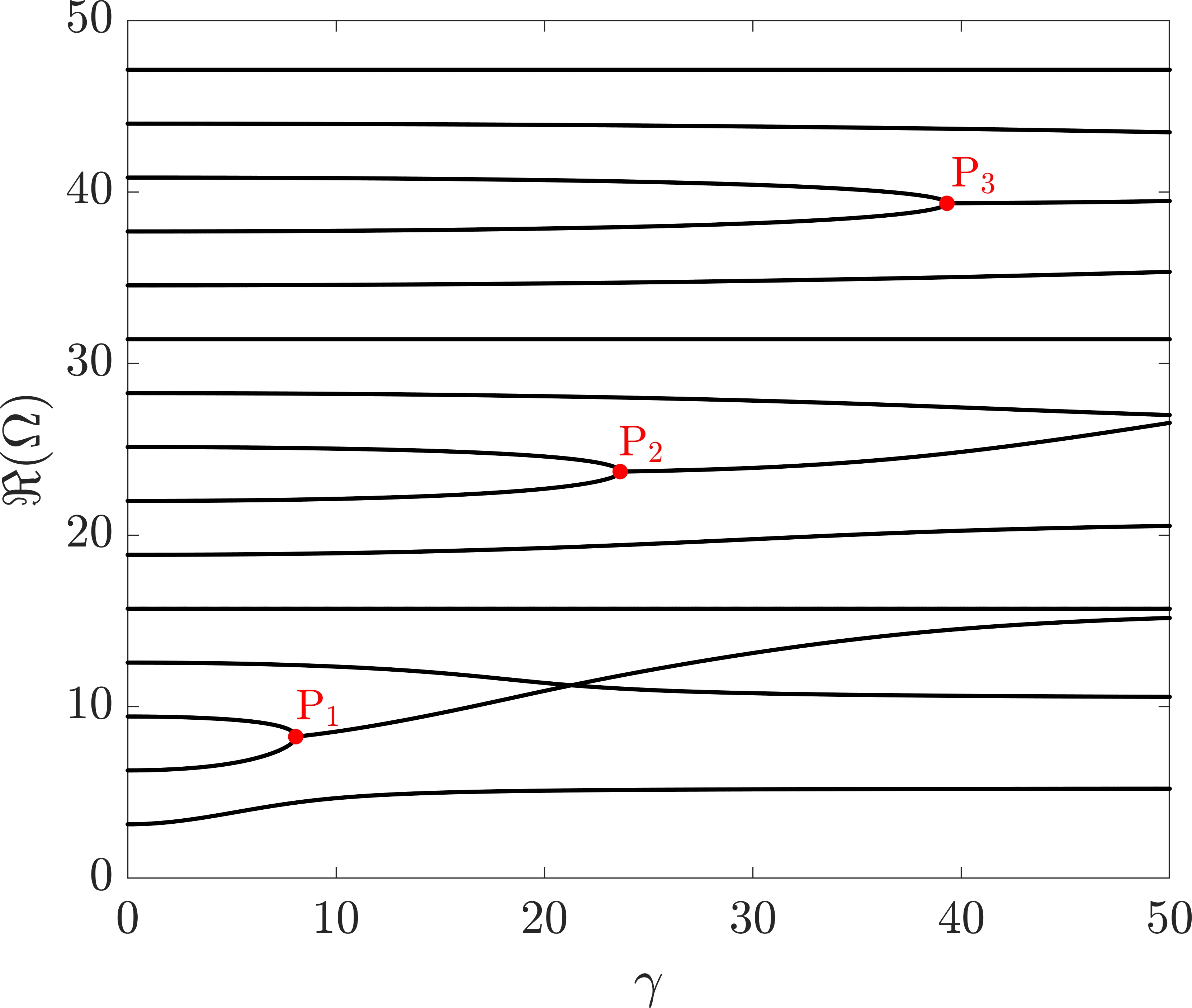}\label{Figrods2b}}\hspace{1mm}
	\subfigure[]{\includegraphics[height=0.275\textwidth]{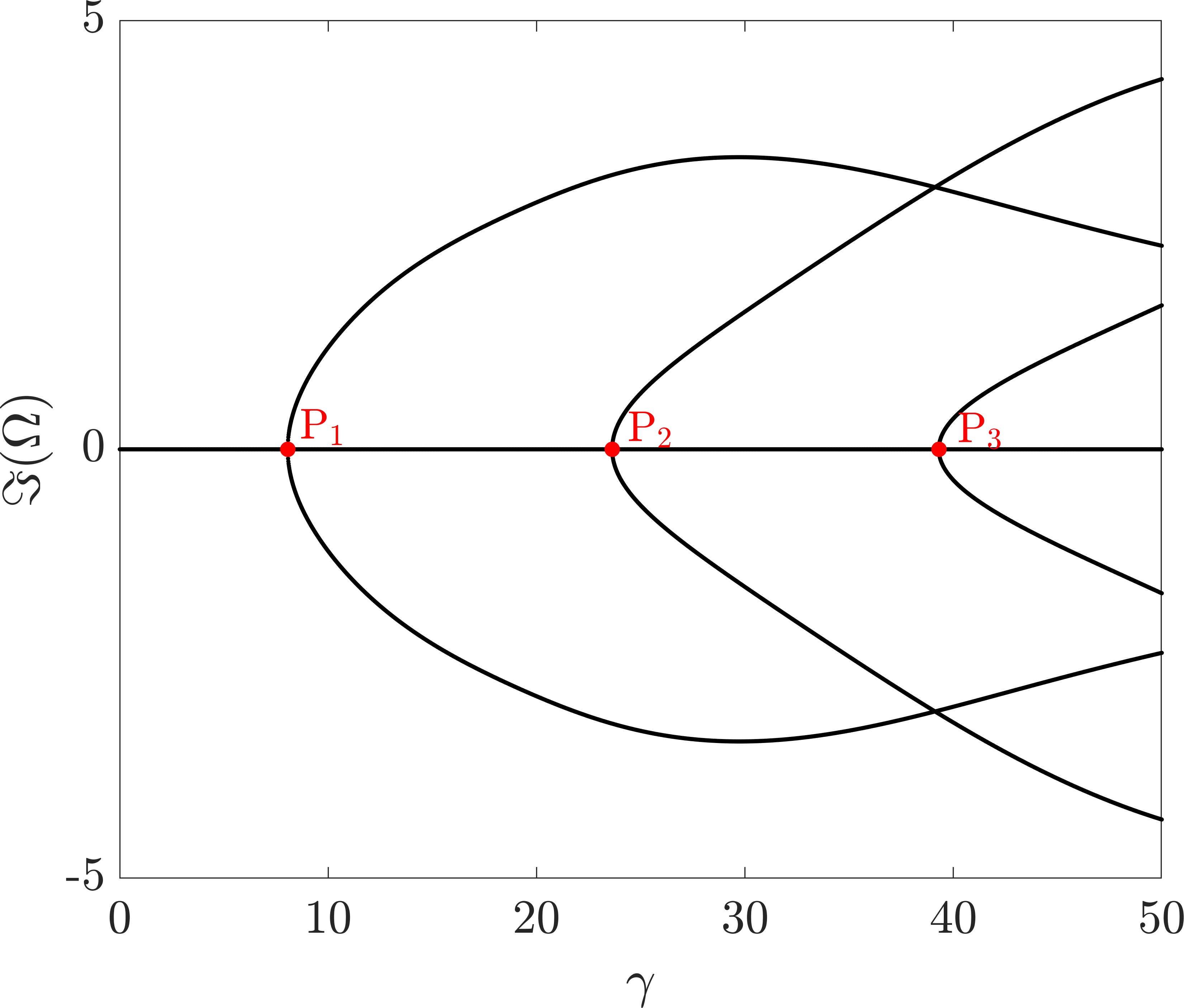}\label{Figrods2c}}
	\caption{Eigenfrequencies of PT-symmetric rod as a function of $\gamma$. The real components for the first 50 modes are displayed in (a) illustrating the presence of several exceptional points. In (b,c),  the real and imaginary eigenfrequency components are respectively displayed zoomed in the shaded blue region of (a).}
	\label{Figrods2}
\end{figure}

Figure~\ref{Figrods2a} displays the real part of the numerically computed eigenfrequencies for the first 50 modes of the PT symmetric rod as a function of $\gamma$. A notable feature is the presence of several exceptional points (EPs) occurring at increasing values of $\gamma$ for modes of increasing order. In Figs.~\ref{Figrods2}(b,c), the real and imaginary eigenfrequency components are respectively displayed zoomed in the region of the first three exceptional points (blue shaded region in Fig.~\ref{Figrods2a}), and are labeled as $P_{1}, P_2$ and $P_3$. Note that these points mark the transition from the region where the two branches forming the EP are purely real, to the region where they become complex conjugate. This is usually referred to as a PT phase transition, from the unbroken (purely real) to the broken (complex conjugate) PT phases~\cite{bender2013observation}. For a continuous medium, the spectrum exhibits several PT phase transitions occurring for different values of $\gamma$. 

\begin{figure}[b!]
	\centering
	\subfigure[]{\includegraphics[height=0.3\textwidth]{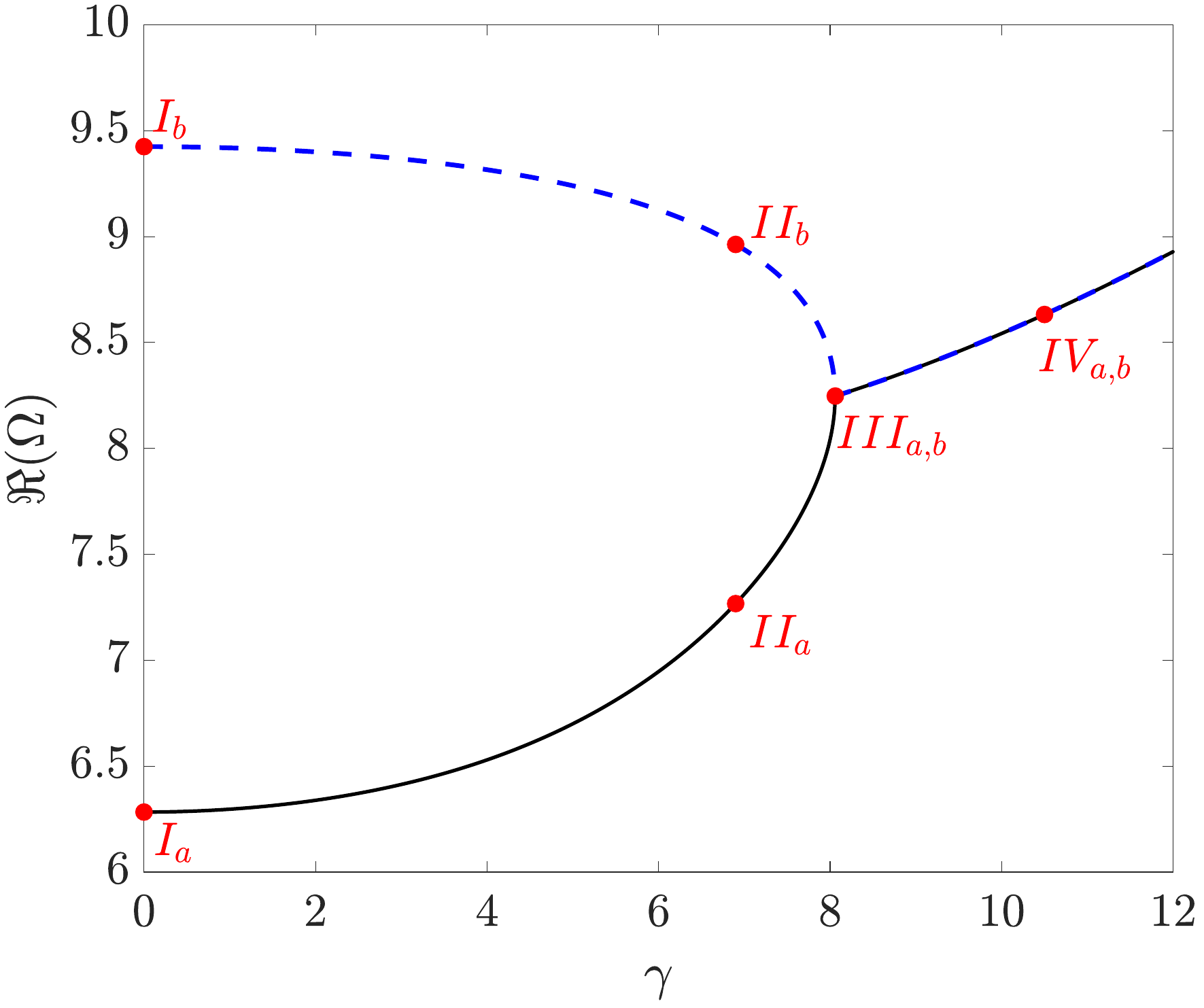}\label{Figrods3a}}
	\subfigure[]{\includegraphics[height=0.3\textwidth]{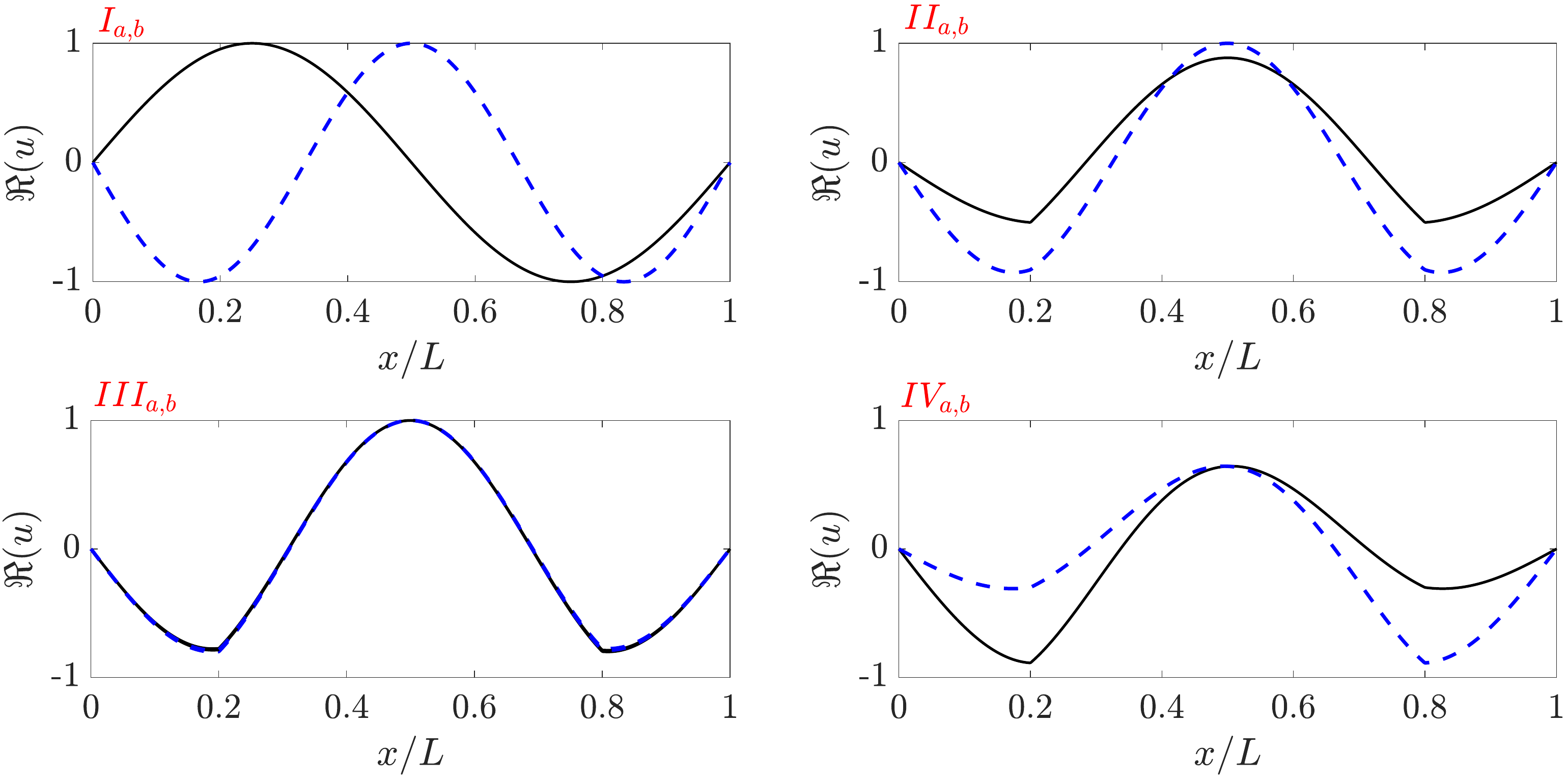}\label{Figrods3b}}
	\subfigure[]{\includegraphics[height=0.3\textwidth]{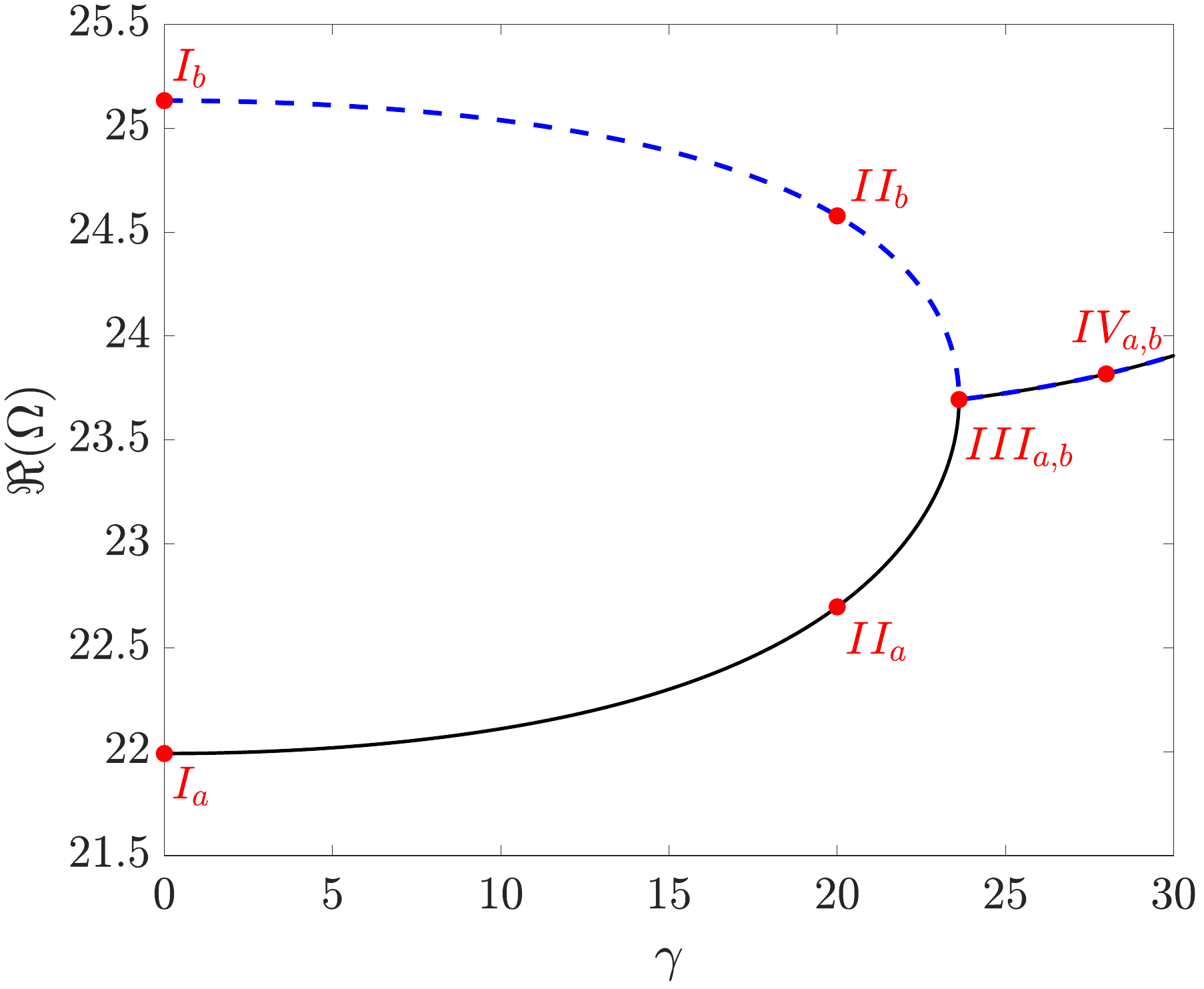}\label{Figrods3c}}
	\subfigure[]{\includegraphics[height=0.3\textwidth]{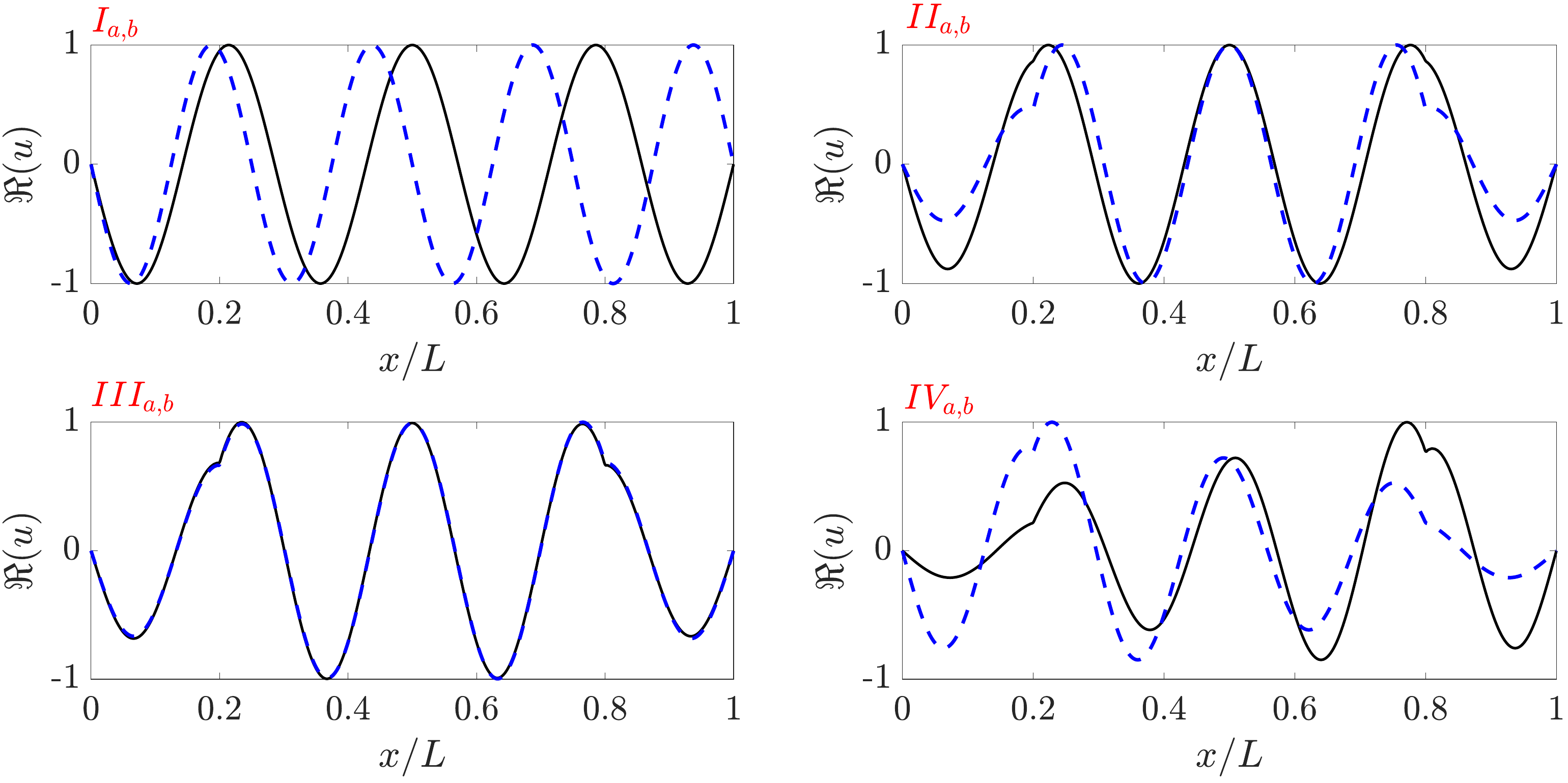}\label{Figrods3d}}
	\caption{Zoomed view on first (a) and second (b) exceptional points of PT symmetric rod, with variation of mode shapes for points labeled in (a,c) displayed in (b,d). Each panel in (b,d) displays two modes whose black solid or blue dashed lines represent solutions corresponding to the branches with same line type in (a,c).}
	\label{Figrods3}
\end{figure}

In Fig.~\ref{Figrods3}, we illustrate the variation of the mode shapes in the vicinity of the first two exceptional points. In particular, Figs.~\ref{Figrods3}(a,c) display zoomed views of the first and second EPs, where solid black and dashed blue lines differentiate the two branches obtained from the numerical solution. Mode shapes corresponding to the points marked in Figs.~\ref{Figrods3}(a,c) are displayed in Figs.~\ref{Figrods3}(b,d), where each panel displays the two mode shapes for a particular $\gamma$, respectively denoted by the solid black and dashed blue lines representing the corresponding line type in (a,c). 
As $\gamma$ increases, the modes hybridize and coalesce at the EP, where, as expected, they become identical. 
After the exceptional point, the two modes seem to be inverted copies of each other with respect to the center of the rod, although we dot not explore this property in further detail.

\begin{figure}[t!]
	\centering
	\subfigure[]{\includegraphics[height=0.23\textwidth]{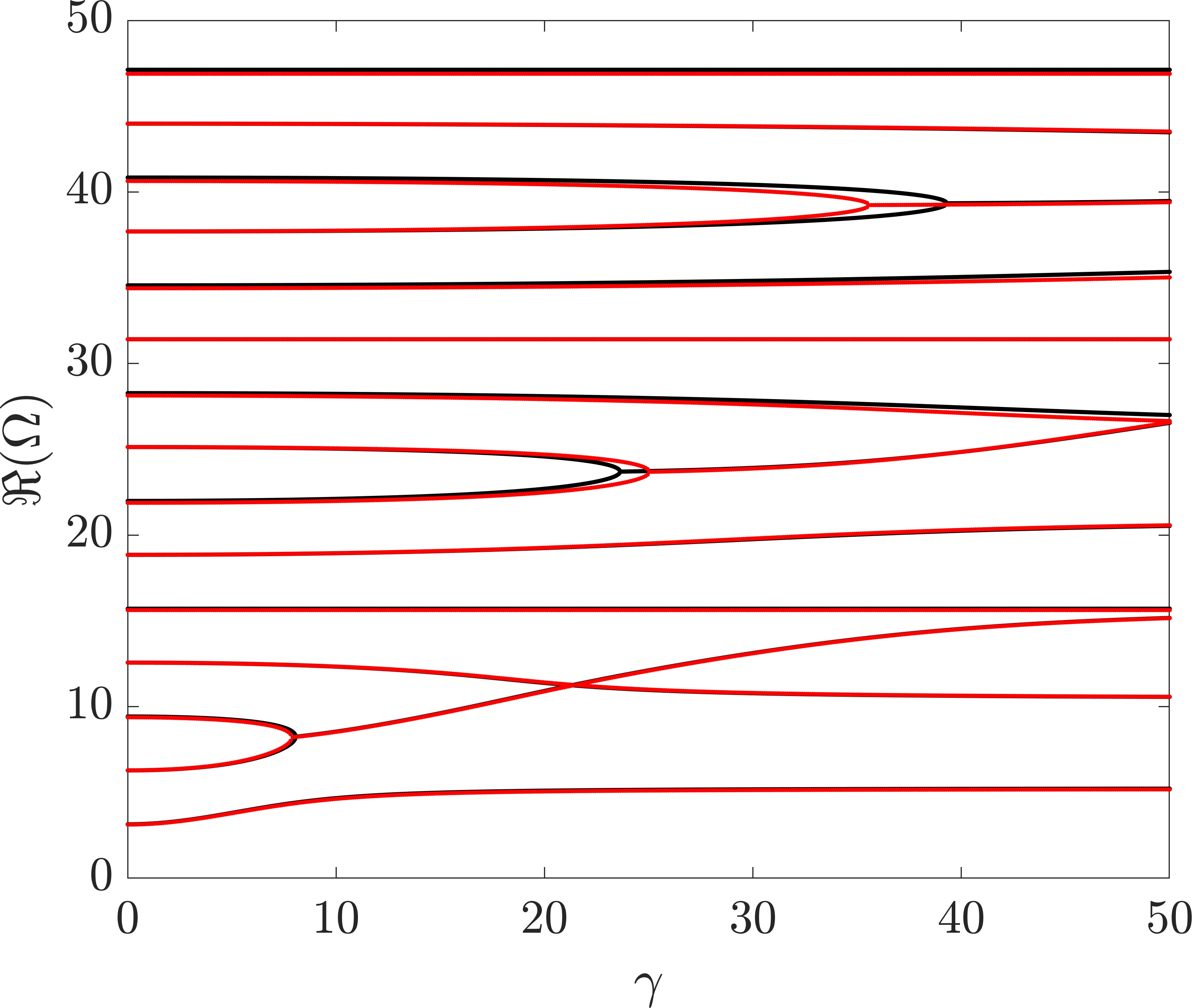}\label{Figrods4a}}
	\subfigure[]{\includegraphics[height=0.25\textwidth]{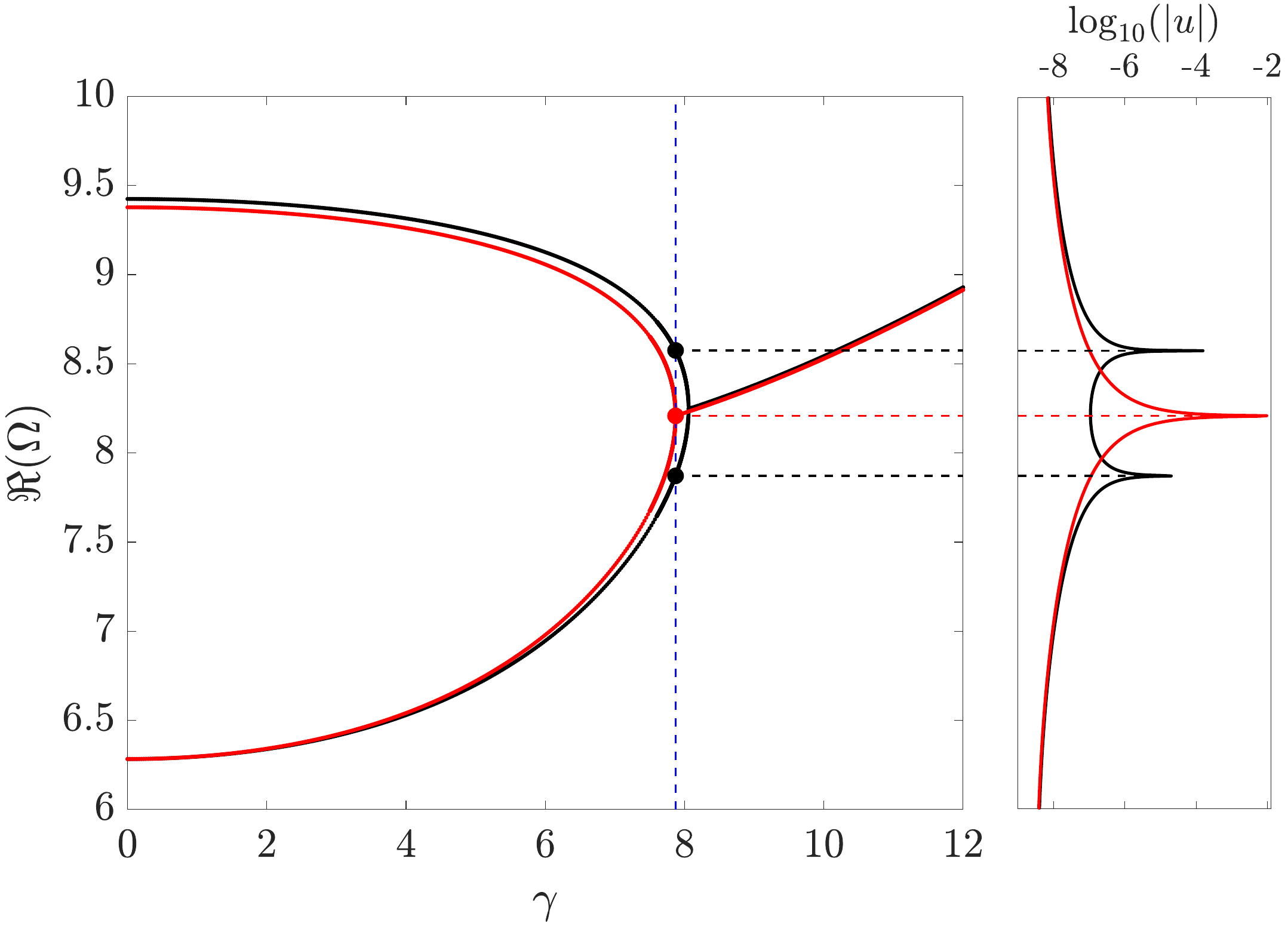}\label{Figrods4b}}
	\subfigure[]{\includegraphics[height=0.25\textwidth]{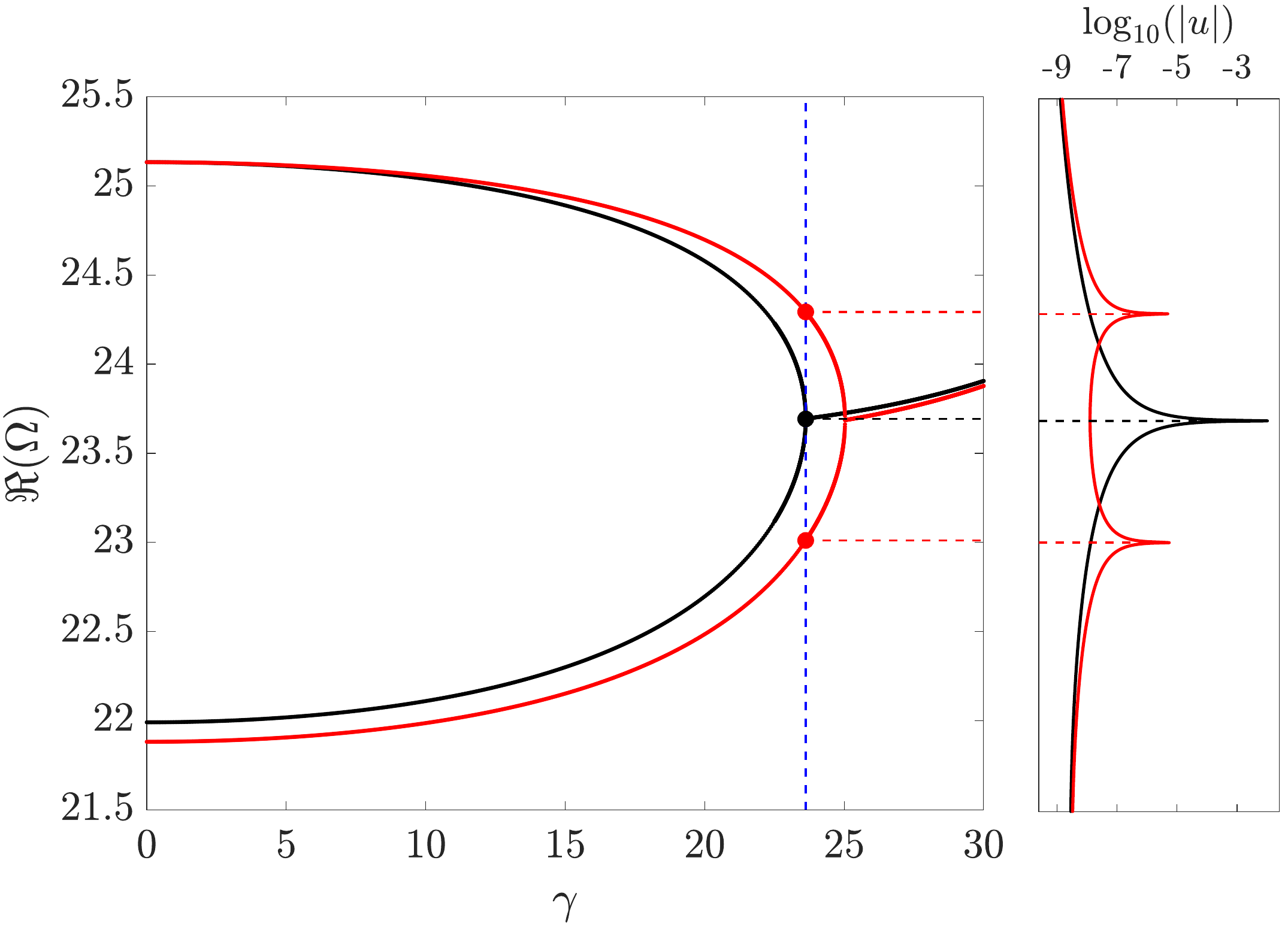}\label{Figrods4c}}
	\caption{Eigenfrequencies of PT symmetric rod without (black) and with (red) point mass inclusion $M_a=0.005\rho AL$ (a), with zoomed views of the first and second exceptional points displayed in (b,c). The right panels in (b,c) display the frequency response of the rod for a fixed $\gamma$ value (dashed blue lines) illustrating the splitting of the resonant peak of the EP into two separate peaks.}
	\label{Figrods4}
\end{figure}

Next, we explore the sensitivity of the EPs to the point mass inclusion $M_a$. In Fig.~\ref{Figrods4a}, we repeat the eigenfrequencies of the PT rod plotted as a function of $\gamma$ (black), superimposed to the frequencies for a mass inclusion $\epsilon=M_a/(\rho A L)=0.5 \%$ (red). This very small inclusion in general introduces small changes to the resonant frequencies; however, larger changes are observed around the EPs. Figures.~\ref{Figrods4}(b,c) display zoomed views of the first and second EPs. In the first case (Fig.~\ref{Figrods4b}), the exceptional point is moved to a lower value of $\gamma$, while in the second case (Fig.~\ref{Figrods4c}) it is moved to a higher $\gamma$. Experimentally, these changes can be detected by measuring the resonant peaks in the frequency response~\cite{hodaei2017enhanced}. This is illustrated in the right panels of Figs.~\ref{Figrods4}(b,c), which display the frequency response at a point $x=0.9L$ resulting from a point force applied at $x_f=0.1L$, obtained from the solution of Eqn.~\eqref{eqforced}. Dashed blue lines are added to identify the fixed $\gamma$ value used in the computation. In Fig.~\ref{Figrods4b}, this line intersects the exceptional point in the added mass case (red dot) and two distinct frequencies on the baseline case with no mass (black dots). The frequency response highlights the split of the resonant peak (red) into two distinct resonant peaks (black) matching the frequencies obtained from the eigenvalue analysis. A similar behavior is demonstrated in Fig.~\ref{Figrods4c}, except that now the dashed blue line intersects the EP for the baseline (no mass) case, and the corresponding resonant peak (black) splits into two peaks (red) upon addition of the attached mass. 

\begin{figure}[b!]
	\centering
	\subfigure[]{\includegraphics[height=0.245\textwidth]{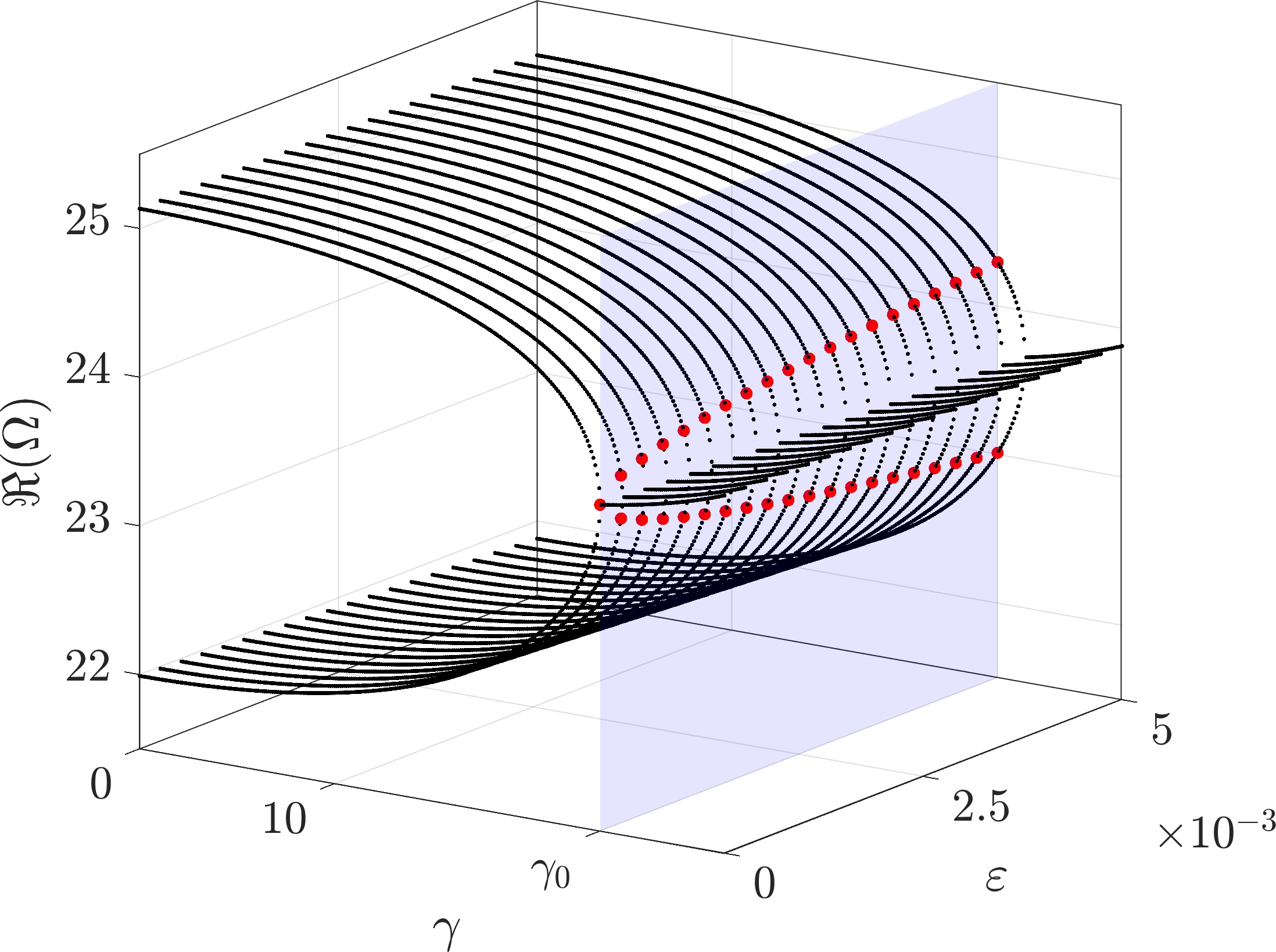}\label{Figrods5a}}
	\subfigure[]{\includegraphics[height=0.245\textwidth]{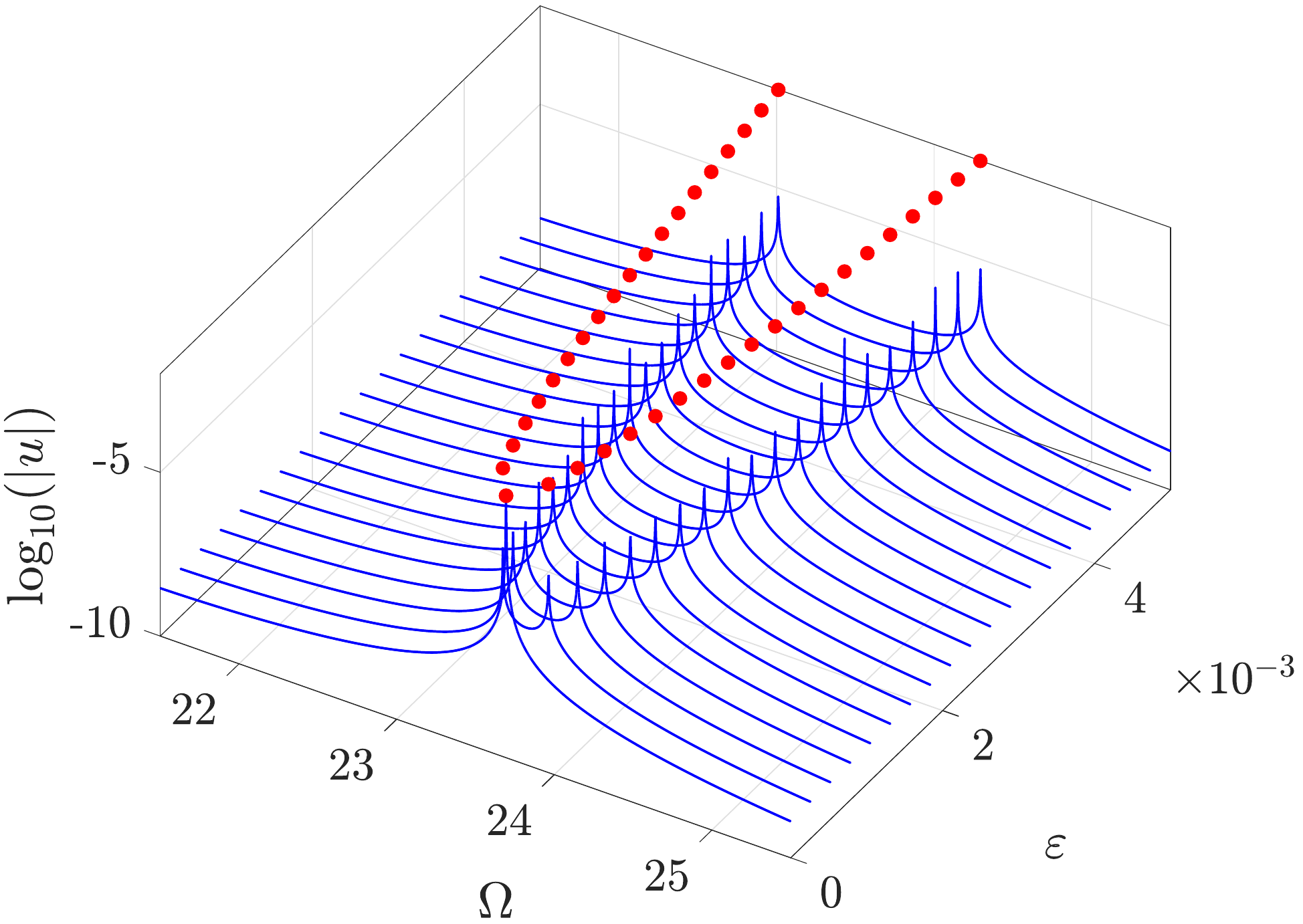}\label{Figrods5b}}\hspace{2mm}
	\subfigure[]{\includegraphics[height=0.245\textwidth]{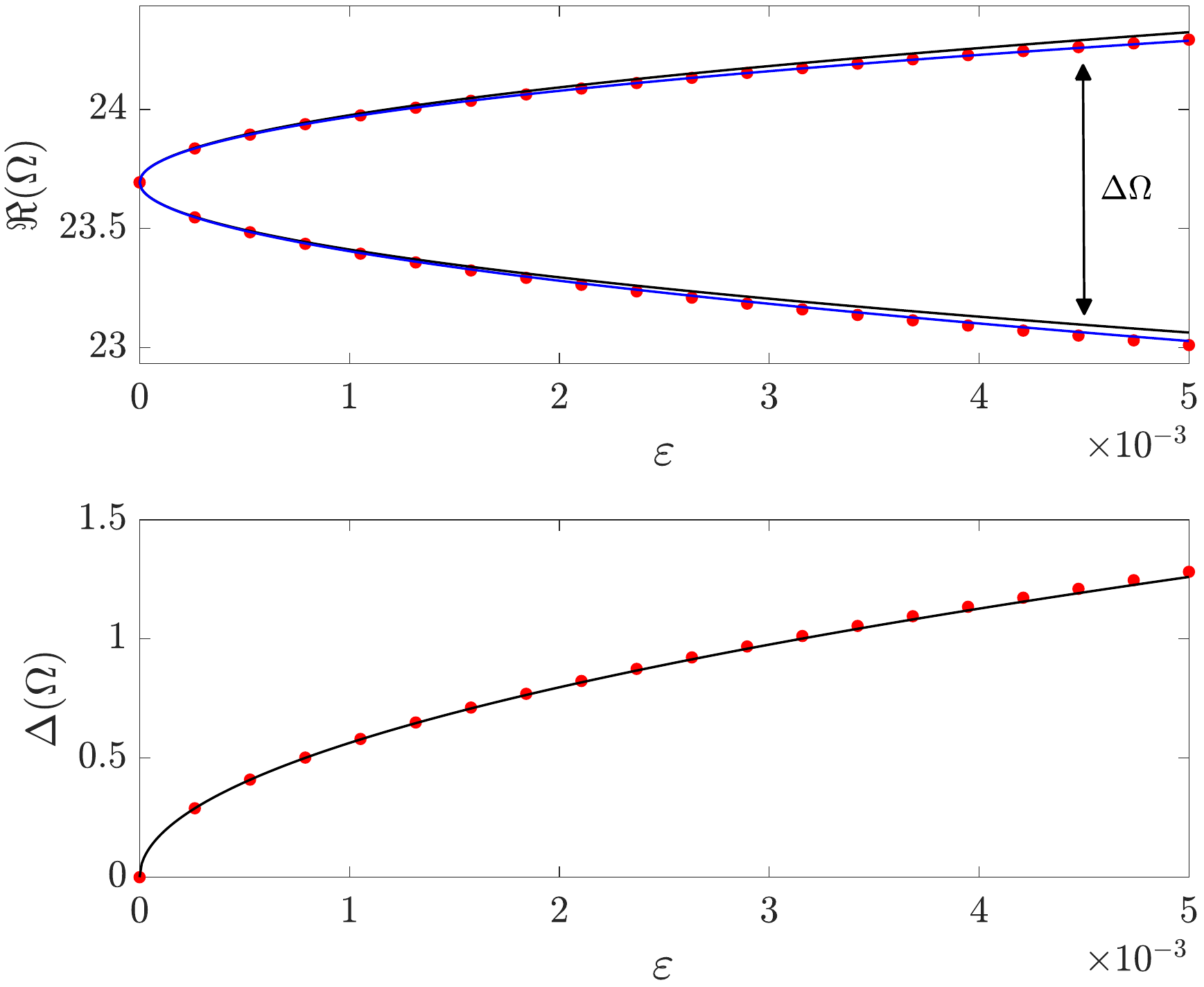}\label{Figrods5c}}
	\caption{Variation of EP as a function of added mass $\epsilon=M_a/(\rho A L)$ (a). The blue plane corresponds to $\gamma=\gamma_0$ defining the EP for $\epsilon=0$, and intersects the frequency plots defining the frequency splitting from the EP (red dots). The frequency splitting can be measured by the variation of the resonant peaks with $\epsilon$ (b). The bifurcation of the EP with $\epsilon$ is repeated in the top panel of (c), where red dots correspond to the numerical solution, while black and blue solid curves correspond to predictions given by the perturbation approach with terms up to $\sqrt{\epsilon}$ and $\epsilon$, respectively. The frequency splitting $\Delta \Omega$ defined by the two branches is displayed in the bottom panel of (c), with the black curve corresponding to the prediction given by the perturbation approach.}
	\label{Figrods5}
\end{figure}

The resonance split around the EPs suggest possible schemes for detection and quantification of the added mass~\cite{hodaei2017enhanced,chen2017exceptional,kononchuk2020orientation,shmuel2020linking}. Figure~\ref{Figrods5a} displays the eigenfrequencies of the PT rod as a function of $\gamma$ for the second exceptional point, for $\epsilon \in [0,\,\,0.5\%]$. The shaded blue plane corresponds to $\gamma=\gamma_0$ defining the EP for $\epsilon=0$, and its intersection with the frequency plots highlight the frequency split as a function of $\epsilon$ (red dots). Indeed, the frequency difference $\Delta\Omega$ corresponding to the split can provide a way to quantify $\epsilon=M_a/(\rho A L)$, as illustrated in Fig.~\ref{Figrods5b}, showing potentials for higher sensitivity due to the $\epsilon^{1/2}$ dependence. For comparison, the eigenfrequencies bifurcating from the EP are repeated in the top panel of Fig.~\ref{Figrods5c}, along with the predictions given by the perturbation approach (solid curves). 
Specifically, the black curve corresponds to the $\epsilon^{1/2}$ approximation, while the blue curve contains also the $\epsilon^1$ dependence. While higher order terms would be required for complete agreement, the perturbation approach produces a good match considering terms up to $\epsilon^1$, and confirms the expected dominant dependence upon $\epsilon^{1/2}$ for $\epsilon<<1$. The frequency splitting $\Delta\Omega$ as a function of the perturbations is often considered as the main sensing parameter~\cite{hodaei2017enhanced}, and is displayed in the bottom panel of Fig.~\ref{Figrods5c}, with the black line corresponding to the prediction given by the perturbation approach (Eqn.~\eqref{eqsplit}). The change in such parameter is theoretically infinite at the onset of the splitting, i.e. $\lim_{\epsilon \to 0} \partial \Delta\omega/\partial \epsilon = \infty$, which is of course limited by the resolution of the frequency measurements~\cite{shmuel2020linking}.

\begin{figure}[t!]
	\centering
	\subfigure[]{\includegraphics[height=0.245\textwidth]{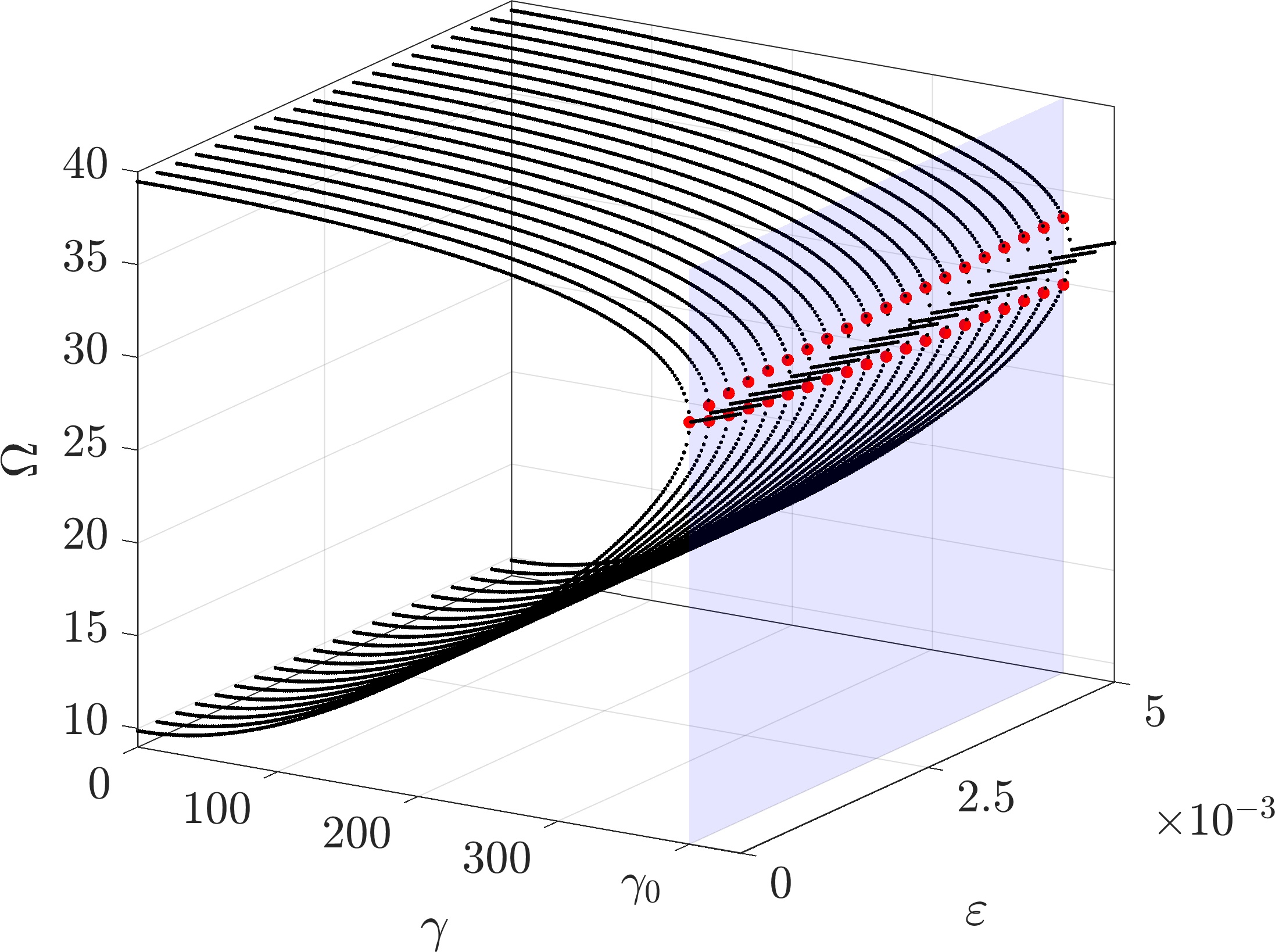}\label{Figbeams1a}}
	\subfigure[]{\includegraphics[height=0.245\textwidth]{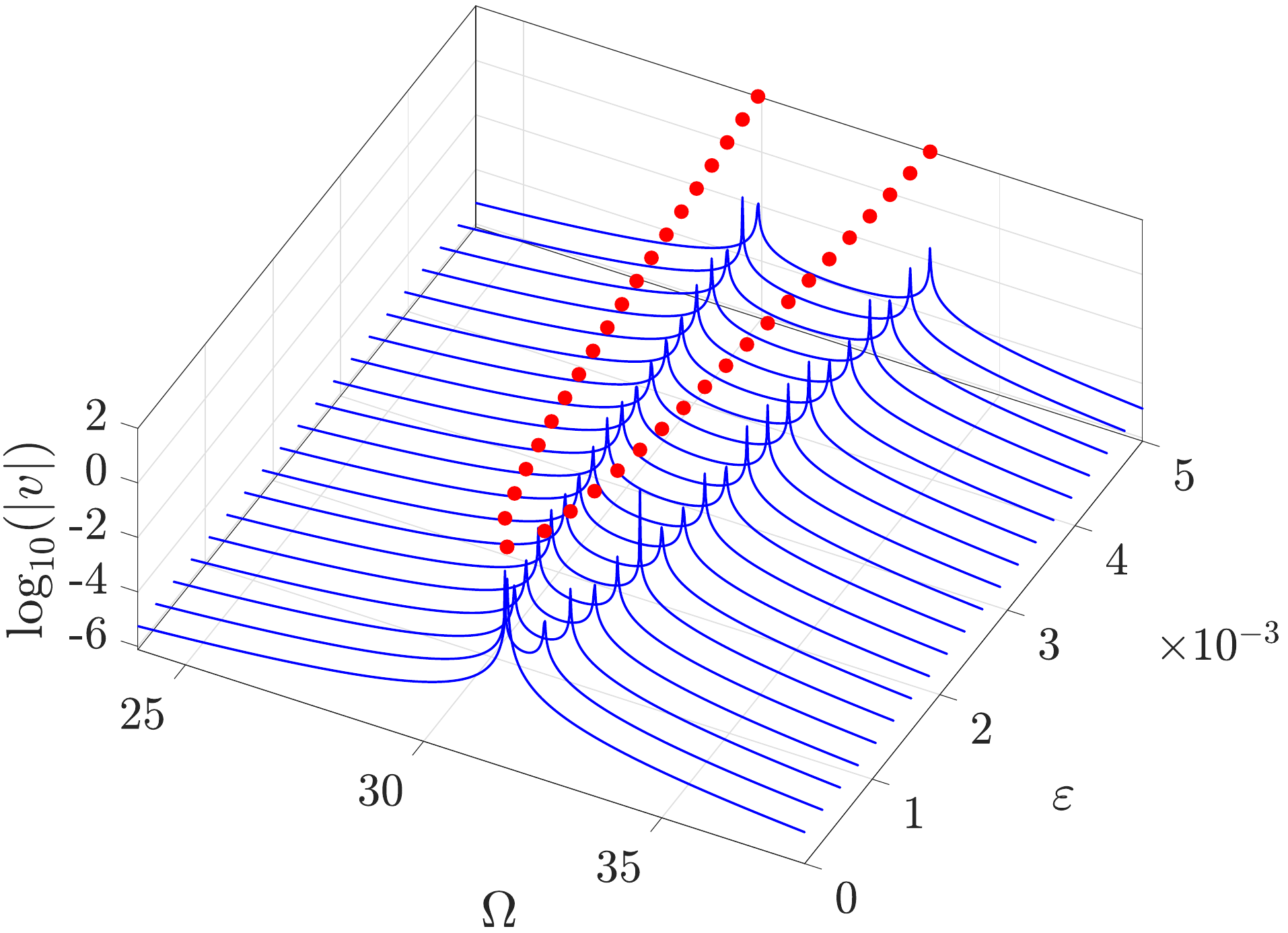}\label{Figbeams1b}}\hspace{2mm}
	\subfigure[]{\includegraphics[height=0.245\textwidth]{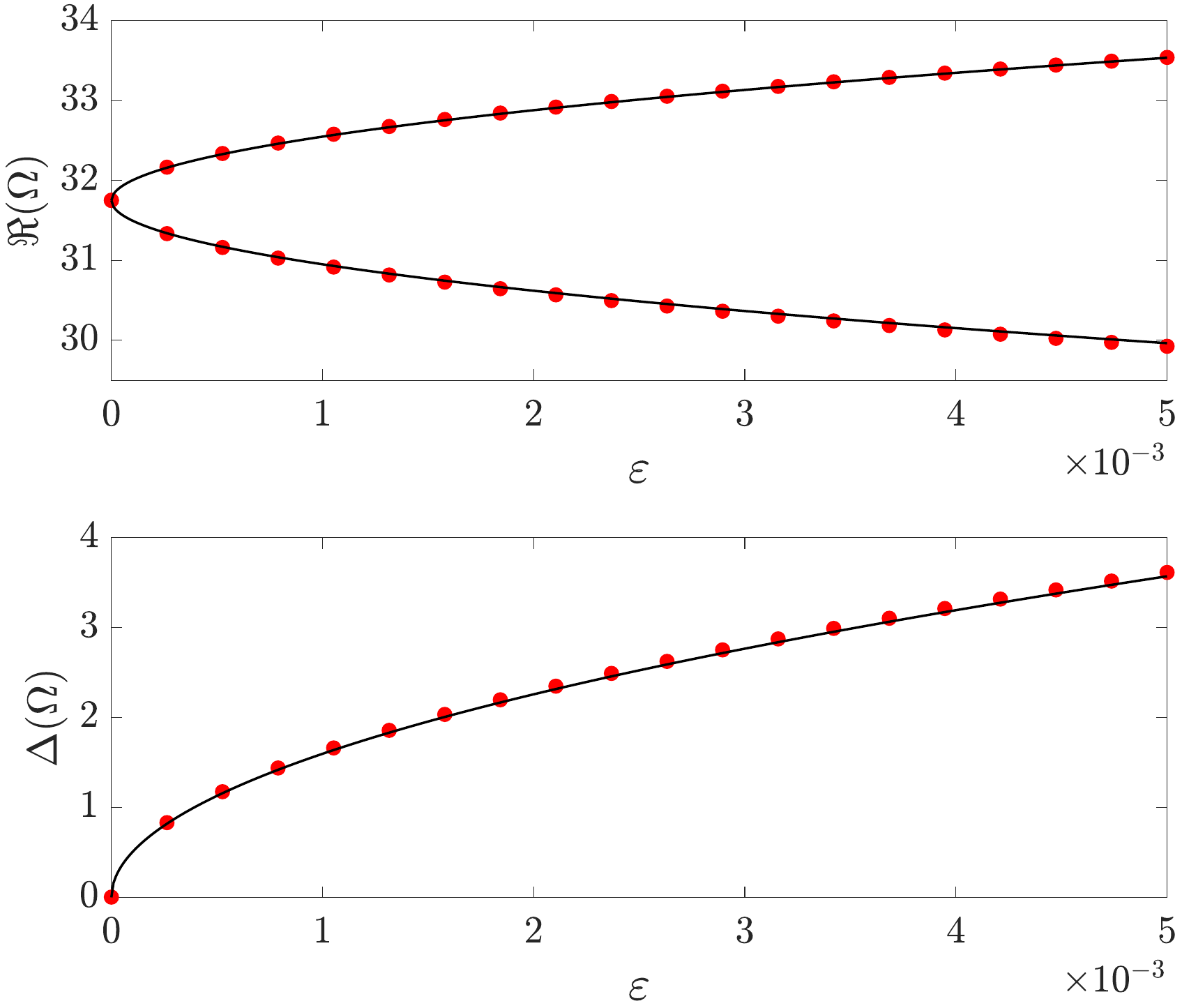}\label{Figbeams1c}}
	\caption{Variation of EP as a function of added mass $\epsilon=M_a/(\rho A L)$ for elastic beam (a). The blue plane corresponds to $\gamma=\gamma_0$ defining the EP for $\epsilon=0$, and intersects the frequency plots defining the frequency splitting from the EP (red dots). The frequency splitting can be measured by the variation of the resonant peaks with $\epsilon$ (b). The bifurcation of the EP with $\epsilon$ is repeated in the top panel of (c), where red dots correspond to the numerical solution, while the black curve corresponds to the prediction given by the perturbation approach illustrating a dependence with $\sqrt{\epsilon}$. The frequency splitting $\Delta \Omega$ defined by the two branches is displayed in the bottom panel of (c), with the black curve also corresponding to the prediction given by the perturbation approach.}
	\label{Figbeams1}
\end{figure}

The presented approach can also be applied to other 1D waveguides such as elastic beams undergoing flexural motion. This is conveniently done by employing Euler-Bernoulli beam theory, where the operator $\mathcal{L}$ defined in Eqn.~\eqref{eqoperator} is replaced by a fourth-order operator. The described numerical procedures are then analogously applied in terms of the vertical displacement $v(x,t)$ of the beam (see \cite{Pal_2019} for more details). Figure~\ref{Figbeams1} displays the results for an elastic beam equipped with the PT symmetric pair of ground springs, where non-dimensional quantities are now defined as $\Omega=\omega/\omega_0$, with $\omega_0=\sqrt{(EI)/\rho A L^4}$, and $\gamma=k/k_0$, with $k_0= EI/(L^3)$, while $I$ denotes second moment of area of the beam cross section. Figure~\ref{Figbeams1a} displays the variation of an exceptional point formed by the first two modes of the beam as a function of $\epsilon$. Similarly to the case of the rods, the EP bifurcates into two branches as a function of $\epsilon$ (red dots), which can be detected by the splitting of resonant peaks in the forced response (Fig.~\ref{Figbeams1b}). The sensitivity of the EP is illustrated in Fig.~\ref{Figbeams1c}, where both numerical results (red dots) and the perturbation analysis results (black curve) confirm the $\epsilon^{1/2}$ dependence.

\section{Guided waves in PT symmetric elastic domains}\label{2Dsec}
The investigations are extended to 2D elastic domains supporting guided waves. Gain and loss are now conceptually introduced through a PT symmetric pair of piezoelectric patches (Fig.~\ref{Figlambschematic}). Piezoelectric transducers are commonly used for the generation of lamb waves\cite{giurgiutiu2007structural,raghavan2005finite,collet2011generation}, for active structural control~\cite{marconi2020experimental,xia2020experimental}, and, more recently, for the investigating the scattering properties of exceptional points~\cite{wu2019asymmetric,hou2018tunable}, to name a few. Therefore, they are excellent candidates for inducing gain and loss through distributions of surfaces stresses that mimick the gain and loss interactions. 

\subsection{Governing equations and transducer modeling}

\begin{figure}[t!]
\includegraphics[width=0.95\textwidth]{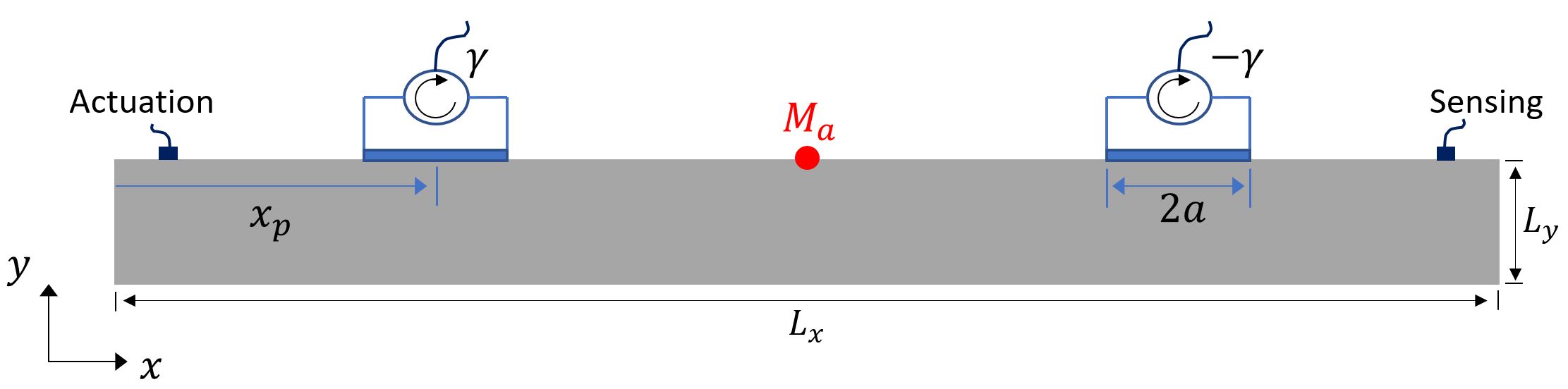}
\centering
\caption{Two-dimensional elastic domain (gray) with PT symmetric pair of piezoelectric patches (blue). Smaller patches (purple) are used for actuation and sensing, and point mass $M_a$ is also attached to the top surface.}
\label{Figlambschematic}
\end{figure}

We consider a rectangular elastic domain in the $x-y$ plane (Fig.~\ref{Figlambschematic}) in plain strain conditions. Unit thickness along the out-of-plane direction is considered for simplicity. The top surface of the domain includes a PT symmetric pair of piezoelectric elements that induce gain and loss, in addition to two piezoelectric transducers used for actuation and sensing. The considered domain includes the point mass $M_a$ also on the top surface at $x=L_x/2$. The equation of motion for the 2D domain is expressed in the frequency domain as~\cite{graff2012wave}
\begin{equation}\label{Eqgov2D}
\omega^2 \rho(x,y)\mathbf{u} + (\lambda+\mu)\nabla(\nabla\cdot\mathbf{u})+\mu\nabla^2\mathbf{u} = \bm{q}_a(x,y,t)+\bm{q}_{PT}(x,y,t)
\end{equation}
where $\mathbf{u}=[u_x(x,y) \hspace{1mm} u_y(x,y) ]^T$ is the displacement vector field, $\lambda$ and $\mu$ are lam\'e constants, and $\nabla=[\partial/\partial x \hspace{1mm}  \partial/\partial y ]^T$. The density is expressed as $\rho(x,y)=\rho_0 + M_a\delta(x-L_x/2)\delta(y-L_y)$, where $\rho_0$ is the density of the domain material.
Also in Eqn.~\eqref{Eqgov2D}, $\bm{q}_a(x,y)$ and $\bm{q}_{PT}(x,y)$ respectively define the area forces associated with actuation and with the gain-loss transducers. The latter consist of a pair of transducers of length $2a$ deposited on the top surface of the domain, centered at locations $x_{c_1}=x_p$ and $x_{c_2}=L_x-x_p$, respectively. As previously noted, the symmetry with respect to the center is necessary to preserve PT symmetry. Their modeling follows references~\cite{raghavan2005finite,collet2011generation}, which assumes their dynamics to be decoupled from that of the 2D domain, and considers their action in terms of stress components applied tangentially to the surface. Under the plane strain assumption, the $\bm{q}_{PT}(x,y,t)=[\phi(x,y,t), \,\, 0]_{y=L_y}^T$, where
\begin{equation}
\phi(x,y=L_y)=V_{0_1}(t) [\delta(x-x_{c_1}+a)-\delta(x-x_{c_1}-a))-V_{0_2}(t)(\delta(x-x_{c_2}+a)-\delta(x-x_{c_2}-a))].
\end{equation}

Here, the applied voltage $V_{0_i}(t)$ is defined by the difference between the $x-$component of the velocities at the edges of the piezo, i.e. $V_{0_i}(t)=\gamma(\dot{u}_x(x_{c_i}+a)-\dot{u}_x(x_{c_i}-a))$, which results from the implementation of a feedback derivative scheme with gain $\gamma$ as suggested by the schematics of Fig.~\ref{Figlambschematic}.  The two additional transducers of length $2d$ and centered at locations $x_a$ and $x_s$ are respectively used for actuation and sensing. According to what previously described, the stress due to the piezoelectric actuation leads to the following expression for $\bm{q}_a=[q_x, \,\, 0^T]$, where:
\begin{equation}
q_x(x,y=L_y,t)=q_0(t) \delta(y-L_y)(\delta(x-x_a+d)-\delta(x-x_a-d)), \qquad \qquad q_y=0,
\end{equation}
with $q_0(t)$ denoting the forcing time history. For simplicity, we consider excitation of unitary amplitude $q_0=1$ in the frequency domain, and we measure the response at the sensing transducer as the integration of the strain $\epsilon_{xx}$ at the top surface over the extent of the sensor $V_o$~\cite{raghavan2005finite}, \emph{i.e.}
\begin{equation}
V_o=\int_{x_s-d}^{x_s+d} \epsilon_{xx}(x,L_y) dx.
\end{equation}


The resulting equation, expressed in Eqn.~\eqref{Eqgov2D}, is conveniently solved using a finite element discretization of the 2D domain within the COMSOL Multiphysics environment. Specifically, under the assumption of zero external loading ($\bm{q}_a=0$), the discretization leads to a polynomial eigenvalue problem that can be solved for eigenfrequencies and mode shapes. Consideration of the external load  leads to a system that is solved in the frequency domain assuming harmoning forcing $q_{x_a}(\omega)$ at specified frequencies.

\subsection{Exceptional points through hybridized Lamb modes}
The 2D elastic domain (Fig.~\ref{Figlambschematic}) is made of aluminum, with material coefficients  $\lambda=40.38$ Gpa, $\mu=26.92$ Gpa and $\rho=2700$ kg/m$^2$. The dimensions are set to $L_x=30$ cm and $L_y=2$ cm, with the transducers of length $2a=4.5$~cm placed at positions $x_{c1}=7.5$ cm and $x_{c2}=22.5$ cm. Also, the actuation and sensing piezoelectric elements have length $2d=0.3$ cm, and are respectively placed at positions $x_a=1.5$ cm and $x_s=28.5$ cm. 

\begin{figure}[t!]
	\centering
	\subfigure[]{\includegraphics[height=0.23\textwidth]{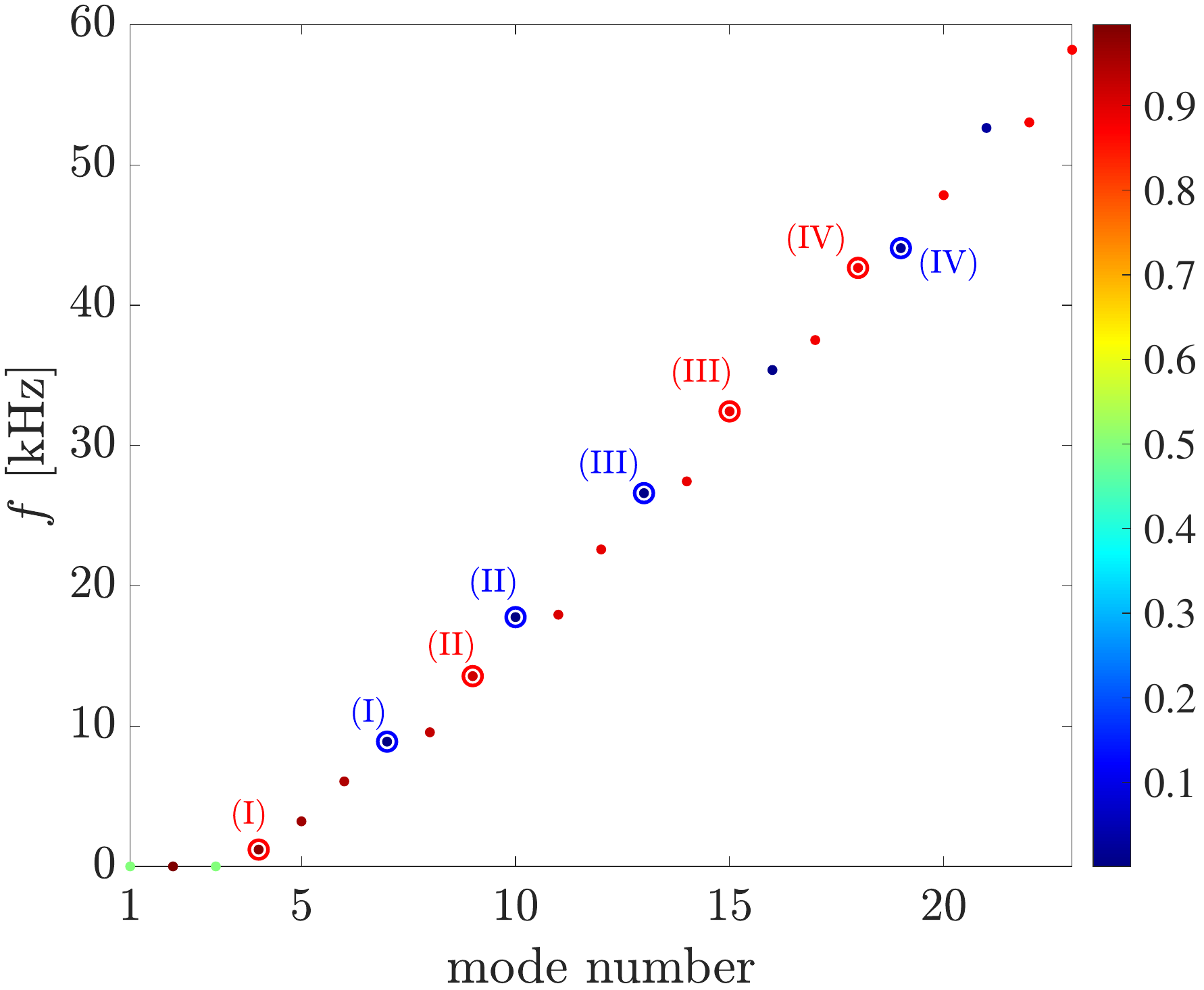}\label{Figlamb1a}}
	\subfigure[]{\includegraphics[height=0.23\textwidth]{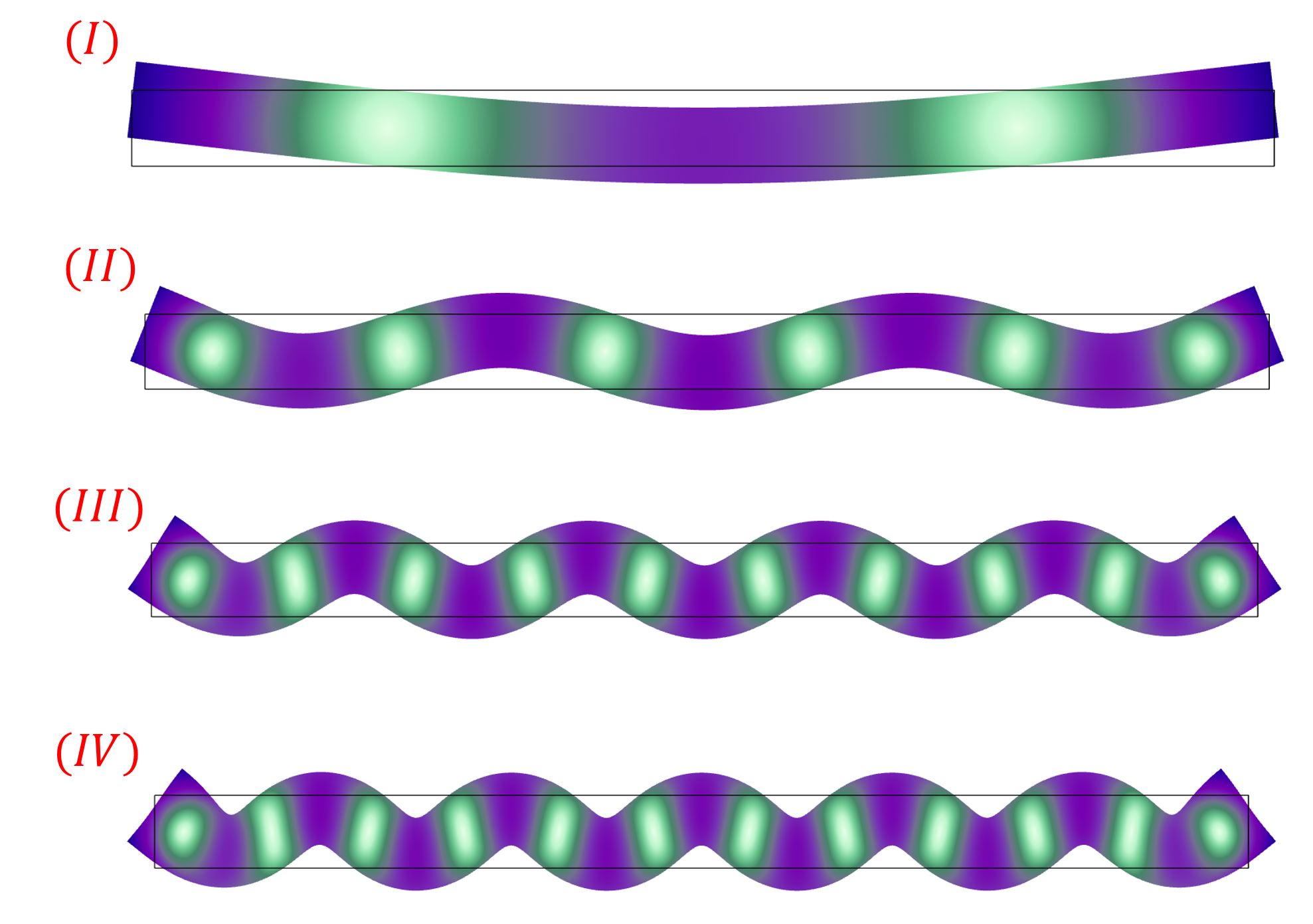}\label{Figlamb1b}}
	\subfigure[]{\includegraphics[height=0.22\textwidth]{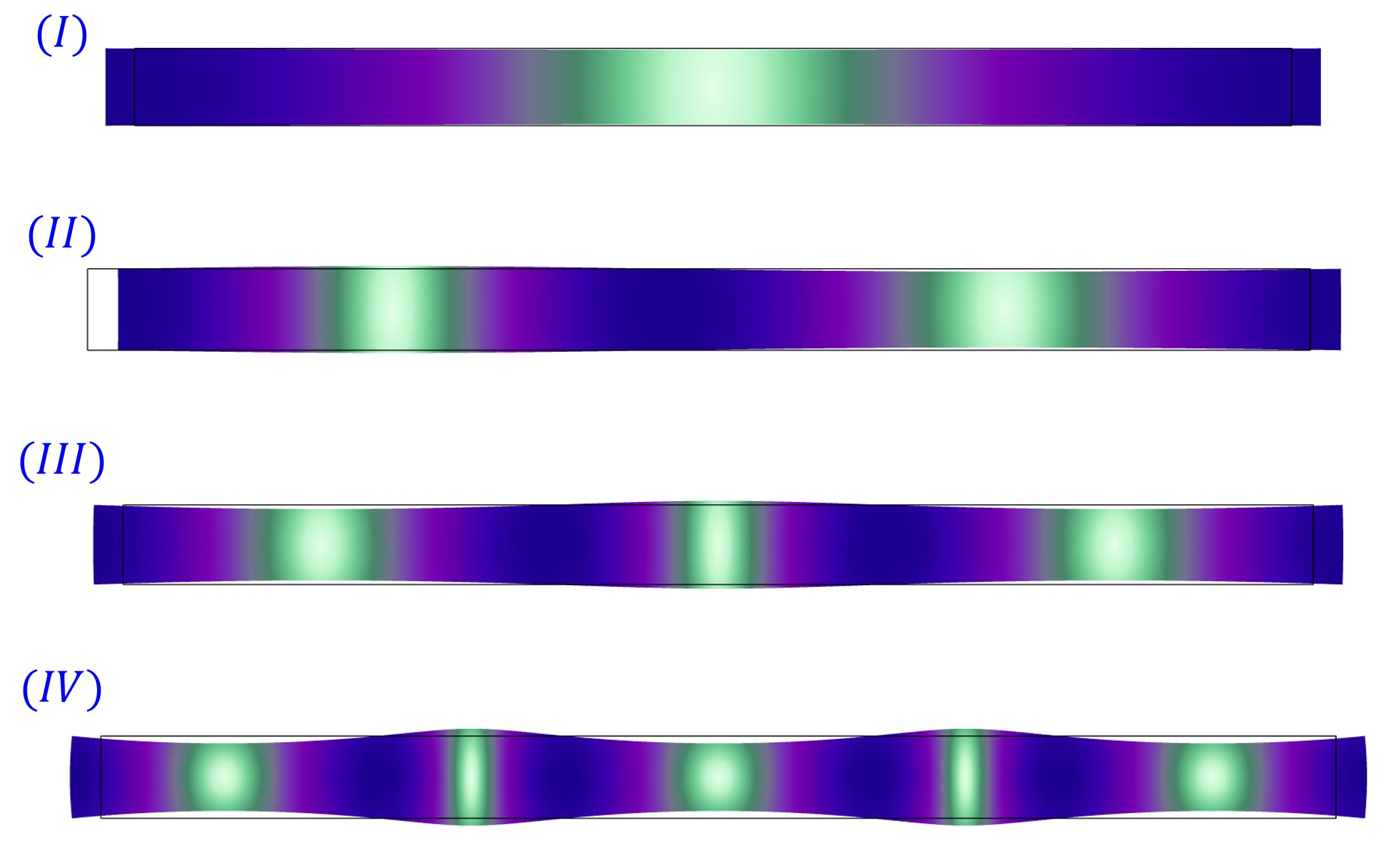}\label{Figlamb1c}}
	\caption{Natural frequencies (a) and deformed mode shapes (b,c) for elastic domain with $\gamma=M_a=0$. The frequencies in (a) are color-coded according to a polarization factor $p$ that identifies modes with predominant flexural (red) or longitudinal (blue) motion.}
	\label{Figlamb1}
\end{figure}

A 2D elastic domain of the dimensions considered supports a family of guided waves (or Lamb waves)~\cite{graff2012wave}, that propagate along the $x$ direction in the form of symmetric (S) or anti-symmetric (A) modes. We consider traction-free boundary conditions in both $x$ and $y$ directions, resulting in a set of vibrating modes that are formed from standing guided waves, as illustrated in Fig.~\ref{Figlamb1}. In particular, Fig.~\ref{Figlamb1a} displays the first natural frequencies of the domain without any influence from the piezoelectric elements or added mass ($\gamma=M_a=0$). Selected modes from the S0 and A0 groups are displayed in Figs.~\ref{Figlamb1}(b,c). These modes are tracked by considering a polarization factor $p$, defined as
\begin{equation}\label{polfaclamb}
p=\dfrac{\int\int u_y^2 dA}{\int\int (u_x^2 + u_y^2) dA}.
\end{equation}
The polarization factor $p$ is employed to color code the scatter plot in Fig.~\ref{Figlamb1a}, where blue dots for $p\rightarrow 0$ correspond to predominantly longitudinal modes ($u_x >> u_y$), while red dots for $p\rightarrow1$ identify predominantly transverse modes ($u_y >> u_x$). This is illustrated in the deformed shapes for selected flexural and longitudinal modes displayed in Figs.~\ref{Figlamb1} (b) and (c), respectively.

\begin{figure}[b!]
	\centering
	\subfigure[]{\includegraphics[height=0.25\textwidth]{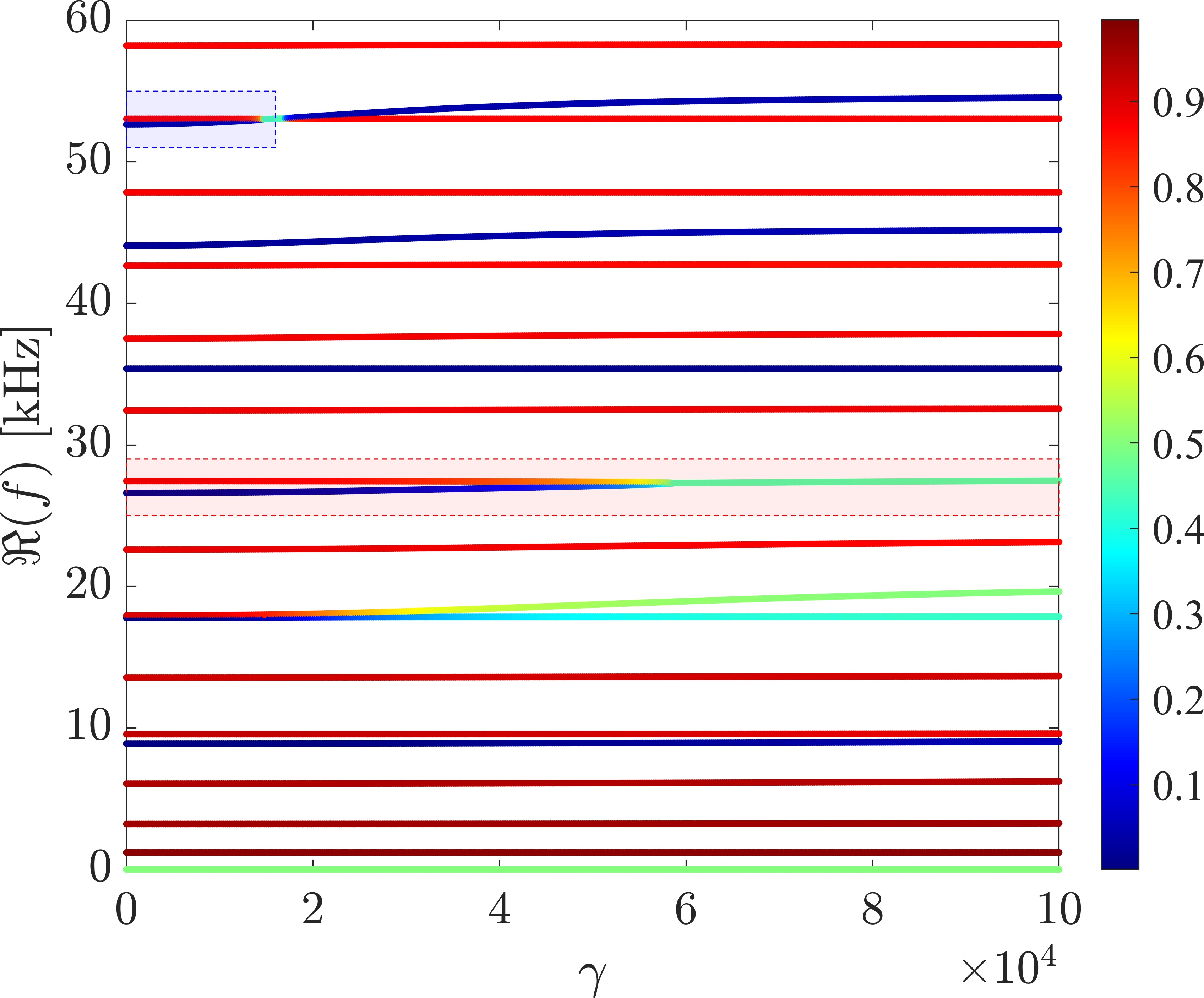}\label{Figlamb2a}}\hspace{2mm}
	\subfigure[]{\includegraphics[height=0.25\textwidth]{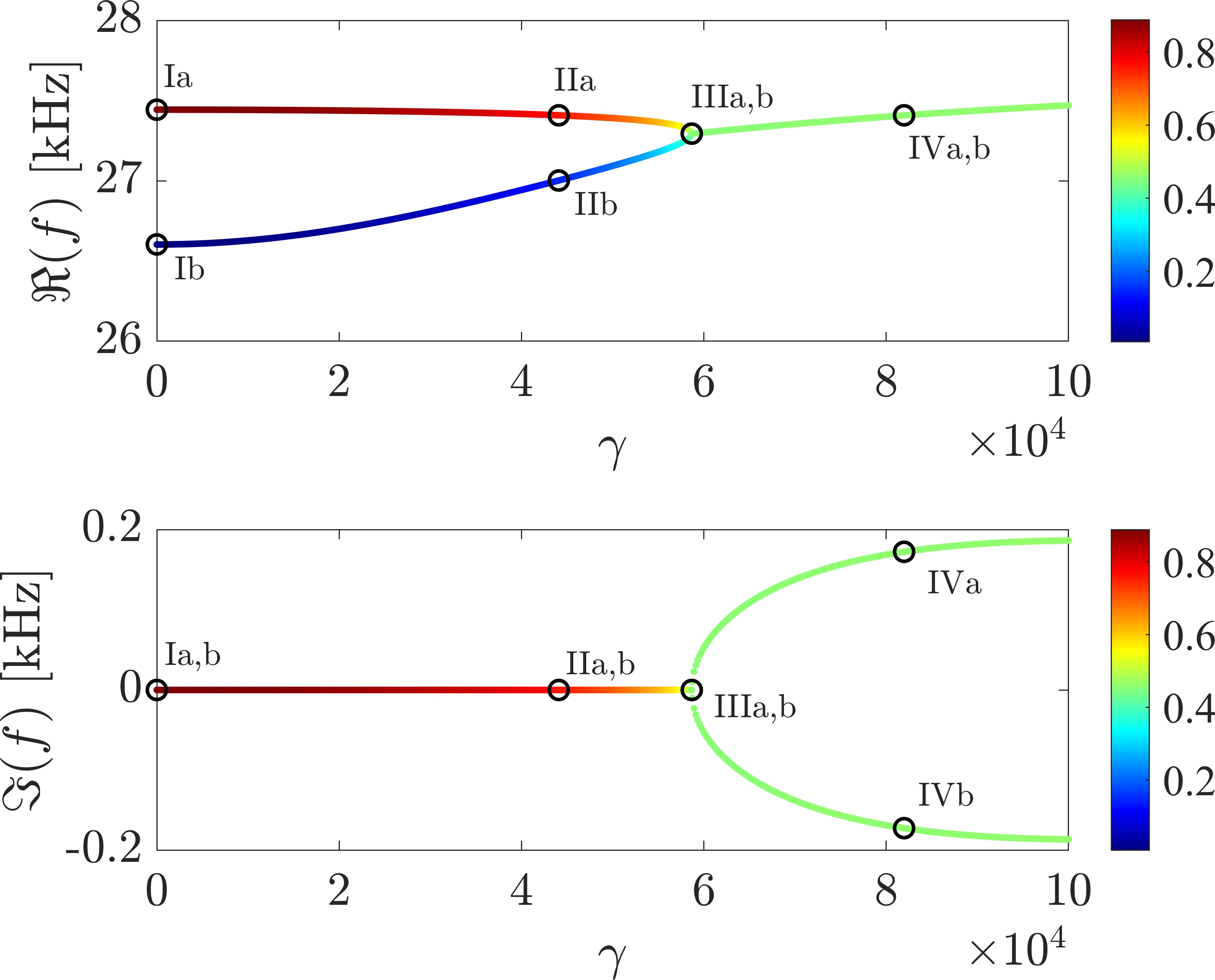}\label{Figlamb2b}}\hspace{2mm}
	\subfigure[]{\includegraphics[height=0.25\textwidth]{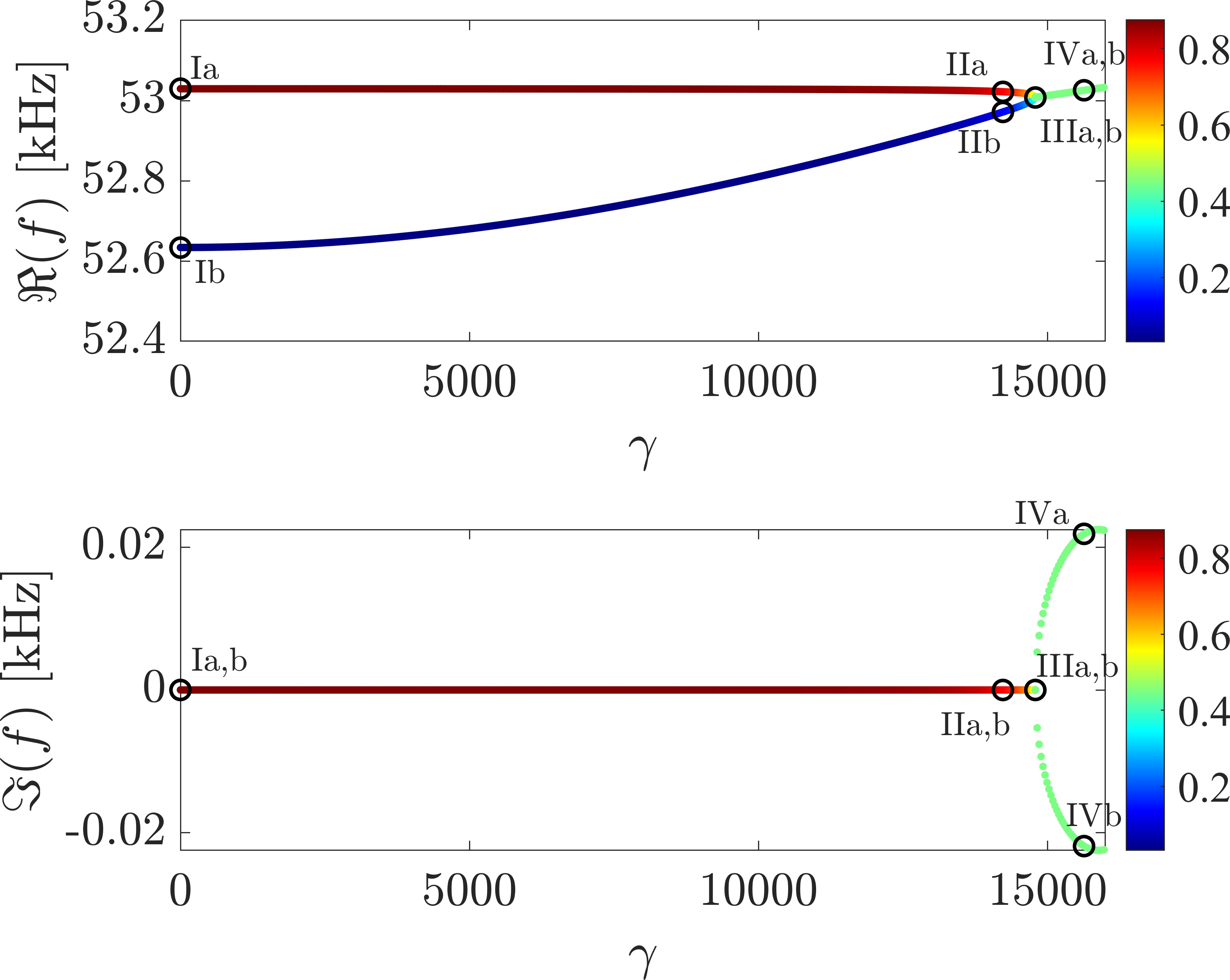}\label{Figlamb2c}}
	\subfigure[]{\includegraphics[width=0.98\textwidth]{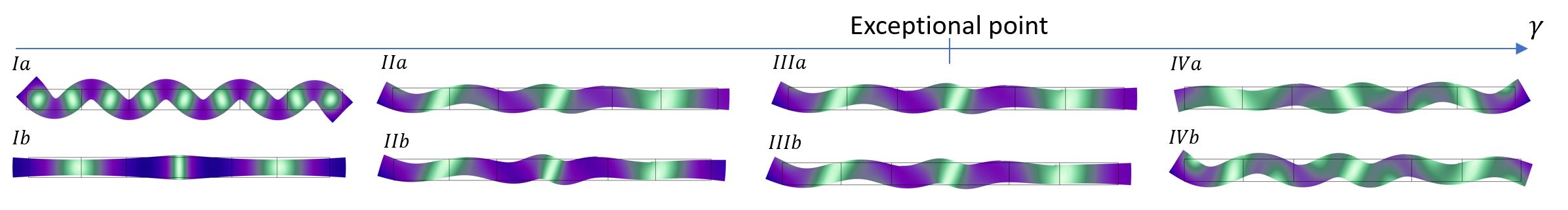}\label{Figlamb2d}}
	\subfigure[]{\includegraphics[width=0.98\textwidth]{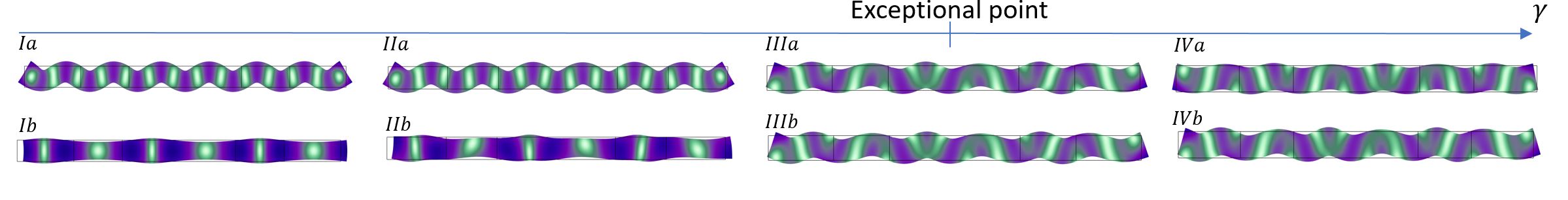}\label{Figlamb2e}}
	\caption{Natural frequencies of 2D domain as a function of $\gamma$ (a), illustrating the formation of two exceptional points. The real and imaginary frequency components of the first and second exceptional points, highlighted by the shaded red and blue areas in (a), are displayed in (b,c). The color change as the two branches in (b,c) merge indicates the hybridization between the longitudinal and flexural modes forming the EP, measured by the polarization factor $p$. The hybridization is further illustrated in (d,e), where deformed shapes for modes marked in (b,c) are displayed.}
	\label{Figlamb2}
\end{figure}

The variation of the corresponding frequencies with $\gamma$ is displayed in Fig.~\ref{Figlamb2a}. The introduction of the PT symmetric pair of piezoelectric transducers leads to the formation of two EPs in the considered frequency range, which are highlighted by the shaded red and blue areas in Fig.~\ref{Figlamb2a}, while Figs.~\ref{Figlamb2}(b,c) display zoomed views of their real and imaginary frequency components. As expected, the EPs define a transition from a region with purely real frequencies, to a region with complex conjugate frequencies. Interestingly, both EPs are formed by the hybridization of a longitudinal (S0) and a flexural (A0) mode, as indicated by the color changes as the branches merge. This color change tracks their change in polarization as quantified by $p$. The deformed mode shapes for selected points marked in Figs.~\ref{Figlamb2}(b,c) are displayed in Figs.~\ref{Figlamb2}(d,e), illustrating the formation of the EP through the hybridization of the modes, with identical mode shapes occurring for the $\gamma$ value corresponding to the EP. 

\begin{figure}[t!]
	\centering
	\subfigure[]{\includegraphics[height=0.33\textwidth]{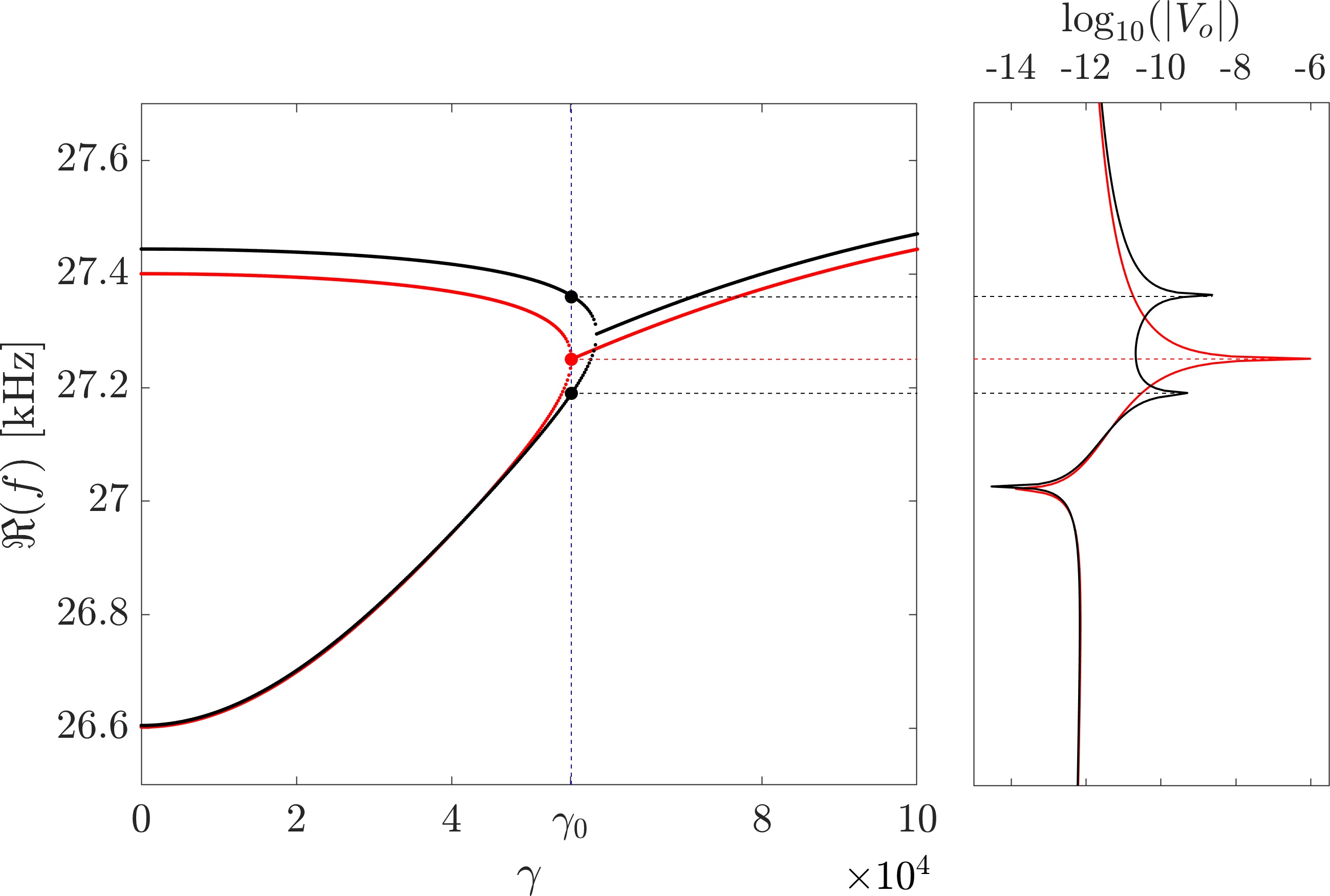}\label{Figlamb3a}}
	\hspace{2mm}
	\subfigure[]{\includegraphics[height=0.33\textwidth]{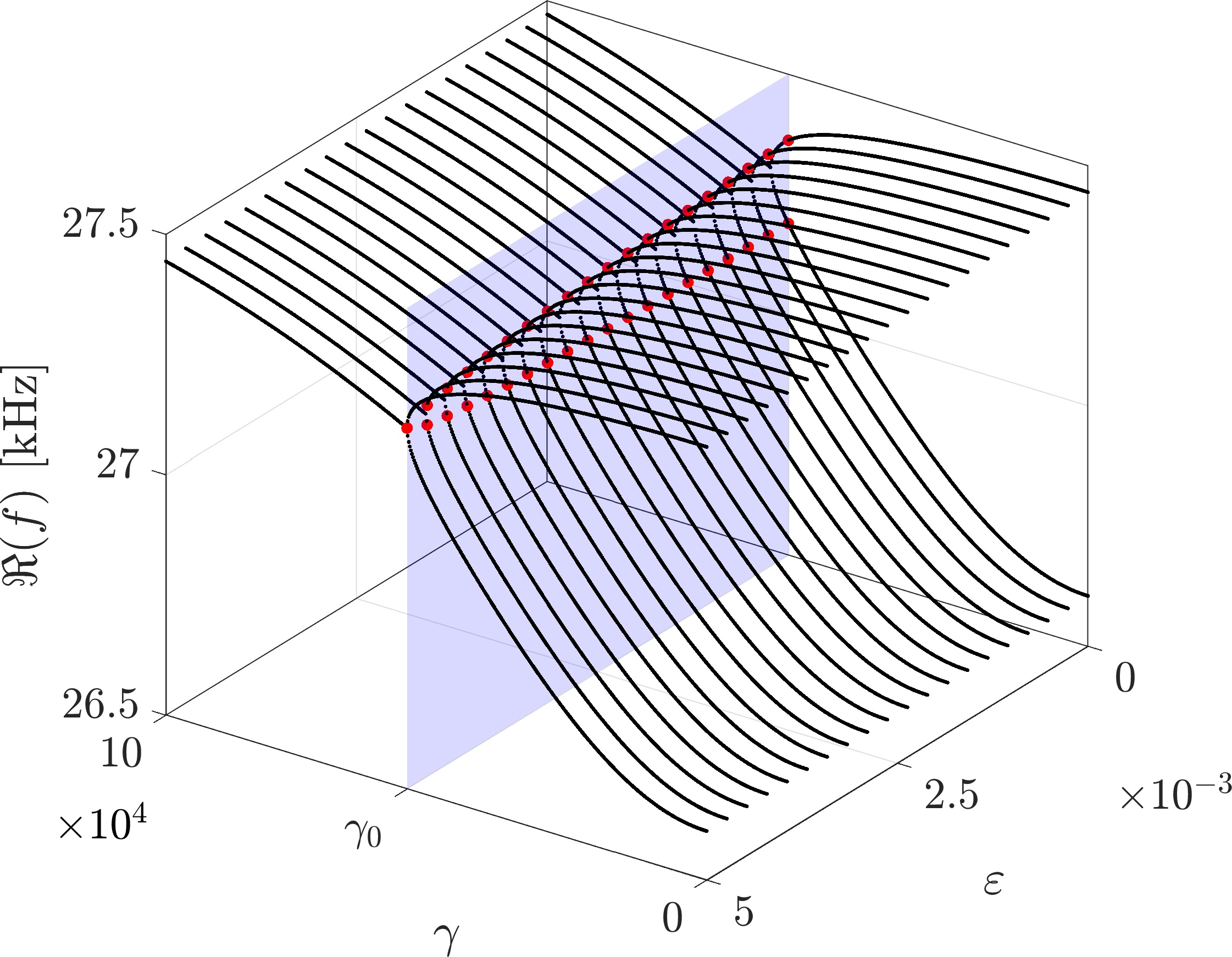}\label{Figlamb3b}}
	\subfigure[]{\includegraphics[height=0.255\textwidth]{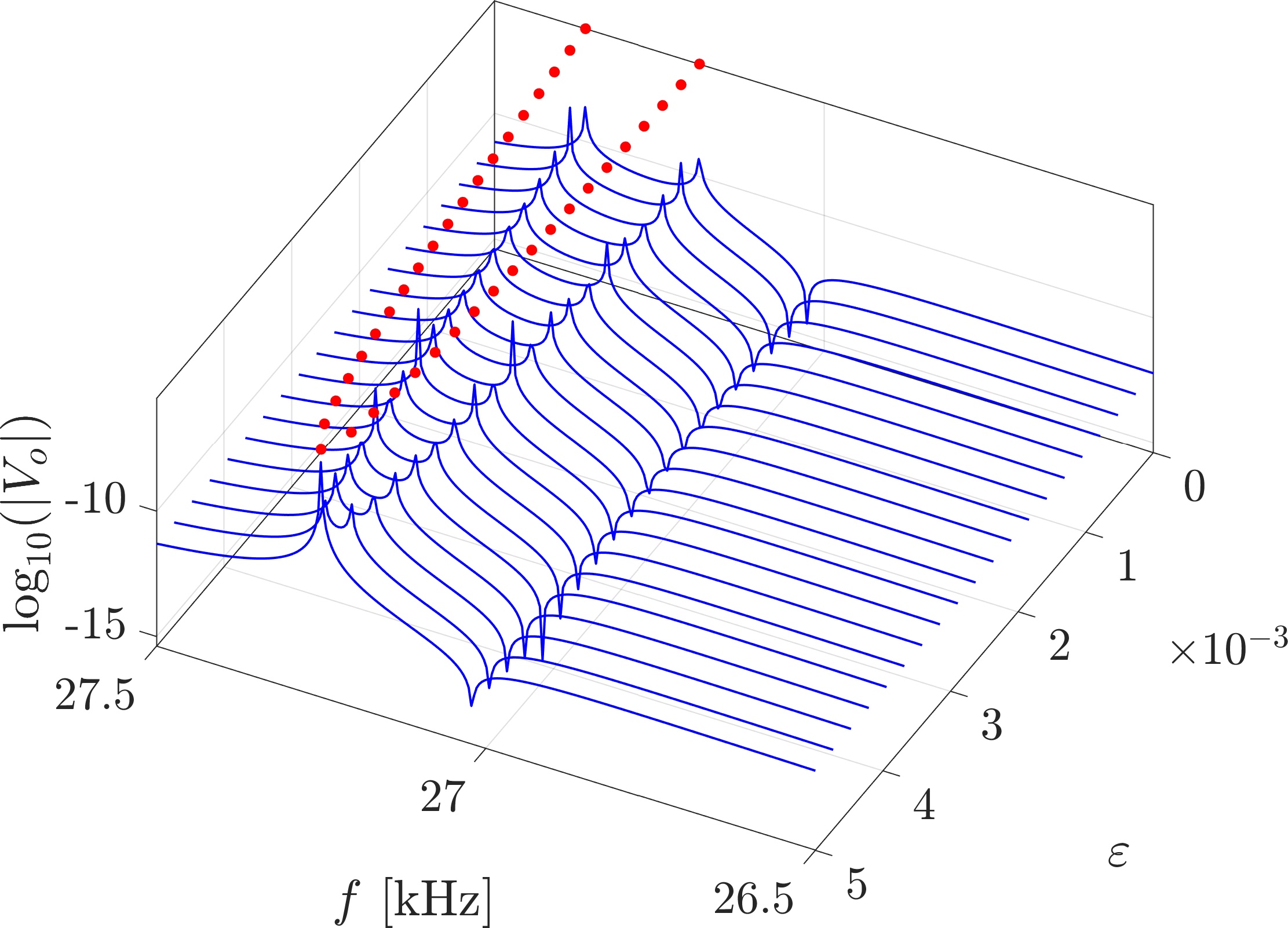}\label{Figlamb3c}}\hspace{1mm}
	\subfigure[]{\includegraphics[height=0.245\textwidth]{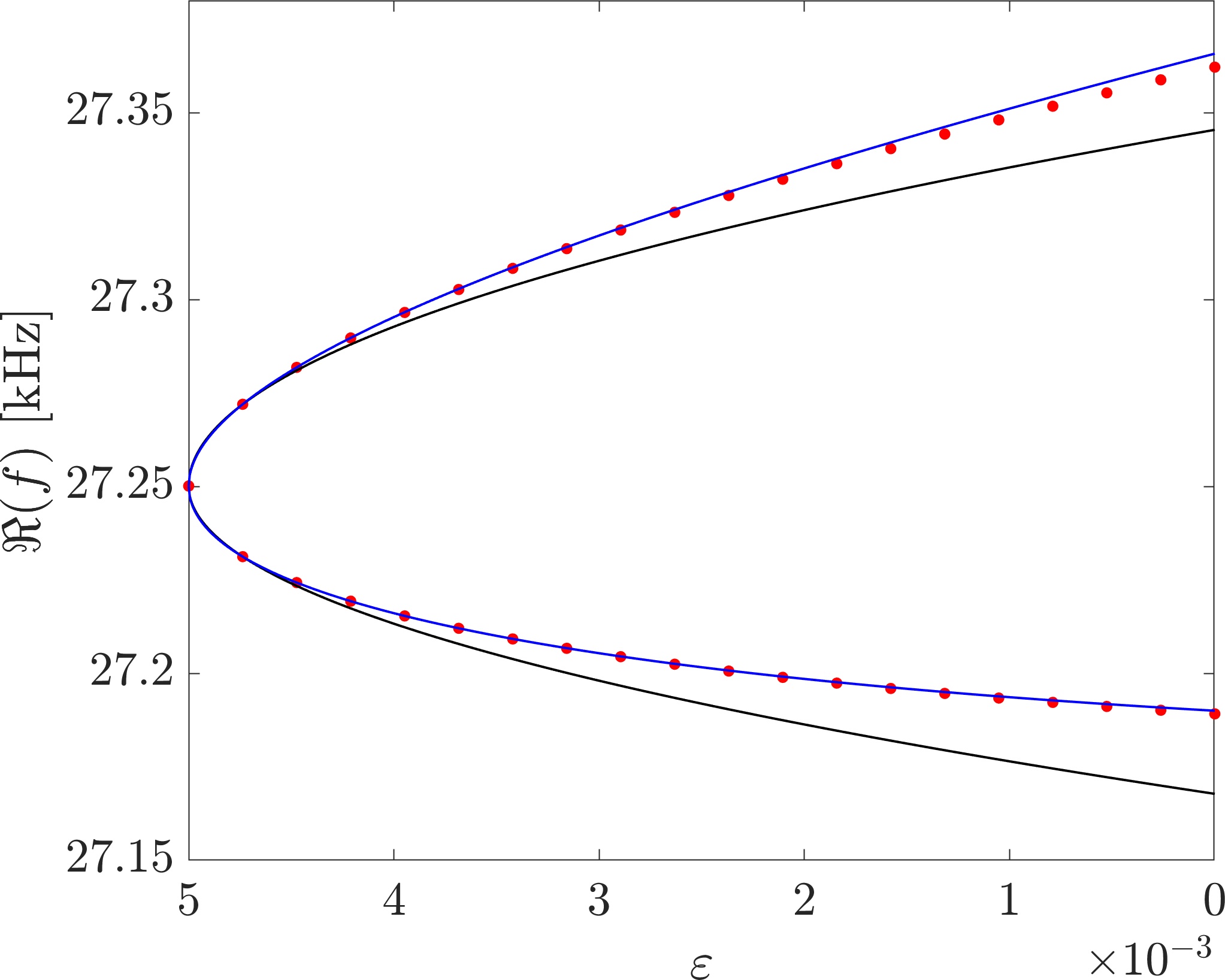}\label{Figlamb3d}}
	\hspace{1mm}
	\subfigure[]{\includegraphics[height=0.245\textwidth]{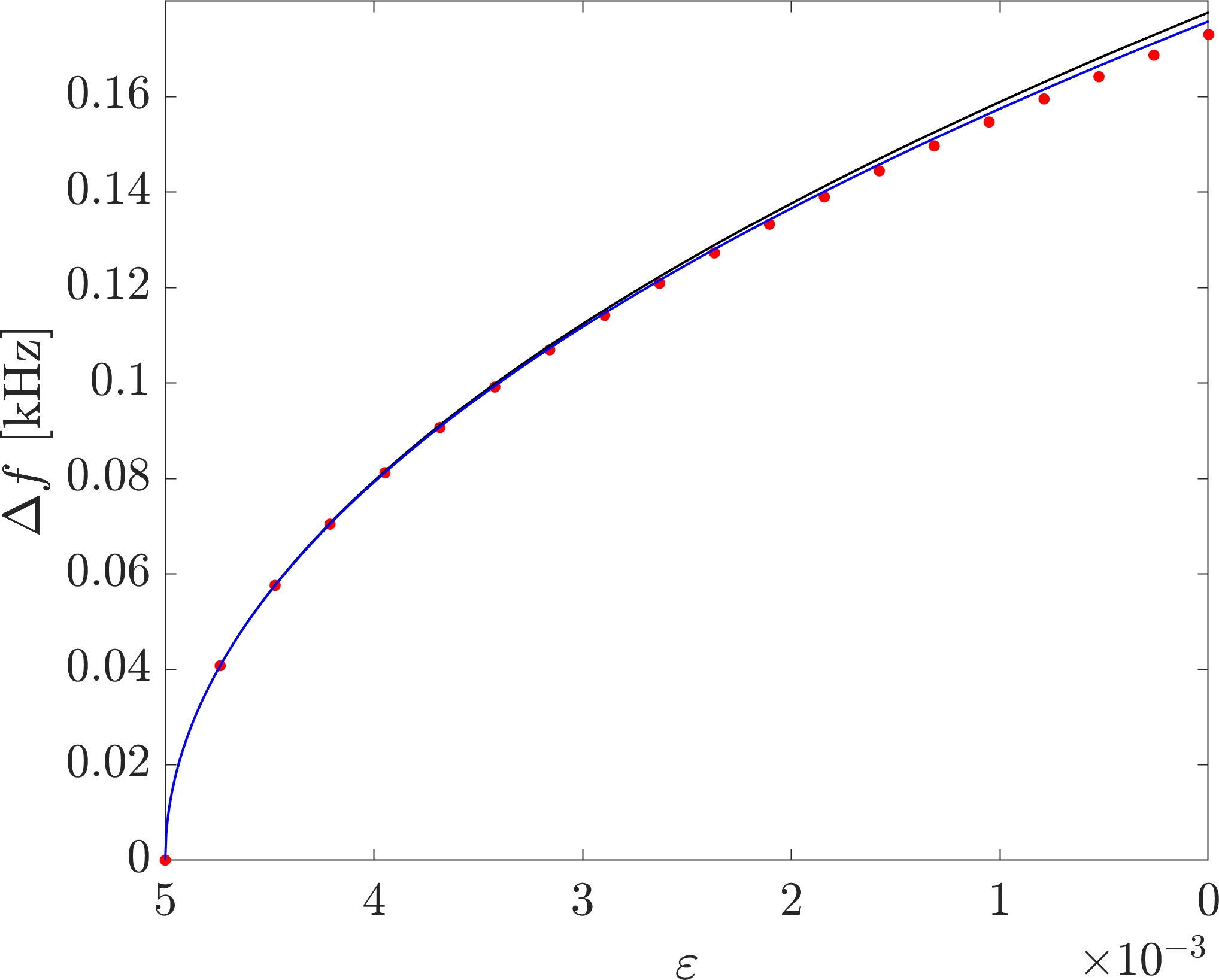}\label{Figlamb3e}}
	\caption{Eigenfrequencies of 2D domain as a function of $\gamma$ with (red) and without (black) added mass $M_a=0.005\rho L_x L_y$ (a). The right panel displays the frequency response measured by the sensor for $\gamma=\gamma_0$, illustrating the splitting of the single (red) into two (black) resonant peaks. The variation of the EP with $\epsilon=M_a/(\rho L_x L_y)$ is displayed in (b), where the blue plane defined for $\gamma_0$ highlights the bifurcation of the EP (red dots). The splitting of the eigenfrequencies with $\epsilon$ is repeated in (c) showing good agreement with the splitting of the resonant peaks measured by the sensor. The frequency splitting is compared to polynomial fits in (d) illustrating the dependence on $\sqrt{\epsilon}$ (black), with better agreement for higher $\epsilon$ values when the linear coefficient is also included (blue). The frequency splitting $\Delta f$ defined by the two branches is displayed in (e), along with polynomial fittings (solid curves) illustrating dominant dependence with $\sqrt{\epsilon}$.}
	\label{Figlamb3}
\end{figure}

Next, we illustrate the sensitivity of the EP to the point mass inclusion. Figure~\ref{Figlamb3a} displays the real frequency component of the first EP for an added mass $\epsilon=M_a/(\rho L_x L_y)=0.5\%$ (red), which is compared to the baseline case with $\epsilon=0$ (black). Similar to the first EP of the rod spectrum (Fig.~\ref{Figrods4b}), the inclusion produces a considerable shift of the EP to a lower $\gamma$ value, which we call $\gamma_0$. On the right panel, the frequency response $V_o$ measured by the sensor illustrates the splitting of the single resonant peak (red) into two separate peaks (black) occurring for $\gamma=\gamma_0$ (dashed blue line). The frequencies of the resonant peaks correspond to the eigenfrequencies for $\gamma=\gamma_0$, as illustrated by dashed red and black lines. Starting from an added mass $\epsilon=0.5\%$, Fig.~\ref{Figlamb3b} illustrates the variation of the eigenfrequencies for decreasing $\epsilon$ values (black), highlighting the bifurcation of the EP into two branches (red dots) for $\gamma=\gamma_0$ (blue plane). The frequency splitting can be measured by the response in the sensor as illustrated in Fig.~\ref{Figlamb3c}, where red dots corresponding to the numerical eigenfrequencies match the frequency of the resonant peaks as $\epsilon$ is varied. Hence, by using the frequency split of the resonant peaks measured by the sensor, one could estimate the value of an inclusion whose mass lies in the considered region. While a perturbation analysis is not conducted for this case, we illustrate in Fig.~\ref{Figlamb3d} that the splitting from the EP occurs according to the ansatz given by Eqn.~\ref{PerturbationExpansions}. In particular, a polynomial of the form $\omega(\epsilon)=\omega_0 + \omega_1\epsilon^{1/2} + \omega_2\epsilon$ can be fitted exactly to the first 3 points of each branch, resulting in the displayed solid curves. The behavior is similar to that of Fig.~\ref{Figrods5b}: the black line corresponds to the solution with only the $\epsilon^{1/2}$ dependence, showing a good agreement for lower $\epsilon$ values, while the blue line considers also the linear correction and better approximates the numerical solution for increasing $\epsilon$ values. The frequency splitting $\Delta f$ is displayed separately in Fig.~\ref{Figlamb3e}, with solid curves corresponding to the polynomial fits. Thus, a sensor operating in such conditions would have a high sensitivity for masses around $\epsilon=0.5\%$, since $\partial \Delta f/ \partial \epsilon \to \infty$ in that region.

\subsection{Exceptional points from surface waves and crack sensing}
\begin{figure}[b!]
\includegraphics[width=0.95\textwidth]{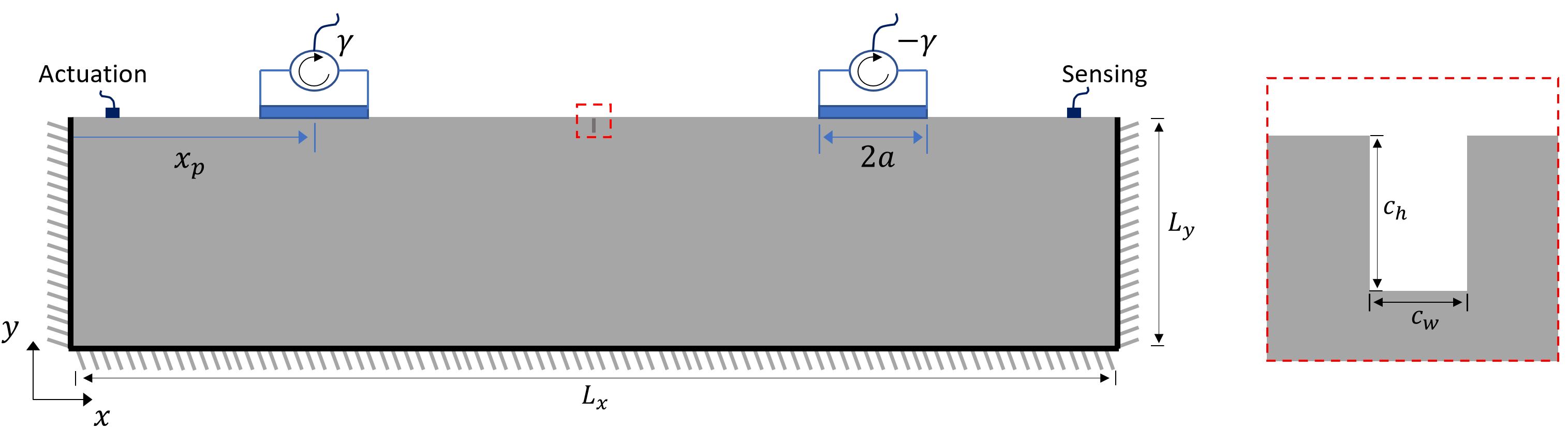}
\centering
\caption{Two-dimensional elastic domain with fixed boundary conditions at all edges except the top surface, where the PT symmetric pair of transducers is deposited. At the center of the top surface, a rectangular crack defect of dimensions $c_w \times c_h$ is illustrated. }
\label{Figsurfschematic}
\end{figure}

We now consider the 2D domain depicted in Fig.~\ref{Figsurfschematic}, similar to that of Fig.~\ref{Figlambschematic} but with a higher height $L_y=9$ cm and fixed boundary conditions on all edges except the top surface. This modification facilitates the formation of vibration modes concentrated at the top surface, where in addition to point mass inclusions we also consider a defect in the form of a rectangular crack of dimensions $c_w \times c_h$. The domain also includes the PT symmetric pair of transducers located at $x_{c1}=7.5$ cm and $x_{c2}=22.5$ cm, with length $2a=3$ cm, and the two smaller elements for actuating and sensing as previously defined. The free surface of the 2D domain supports surface (or Rayleigh) waves~\cite{graff2012wave}, whose wave speed can be approximated as $c_r=c_s(0.87+1.12\nu)/(1+\nu)$, where $c_s=\sqrt{\mu/\rho}$ is the shear wave speed, and $\nu=0.3$ is the Poisson's ratio. This approximation gives $c_r=2929.27$ m/s for the case at hand. We illustrate the formation of exceptional points from such surface wave modes, and their sensitivity to point mass inclusions and to the crack defect.

\begin{figure}[t!]
	\centering
	\subfigure[]{\includegraphics[height=0.35\textwidth]{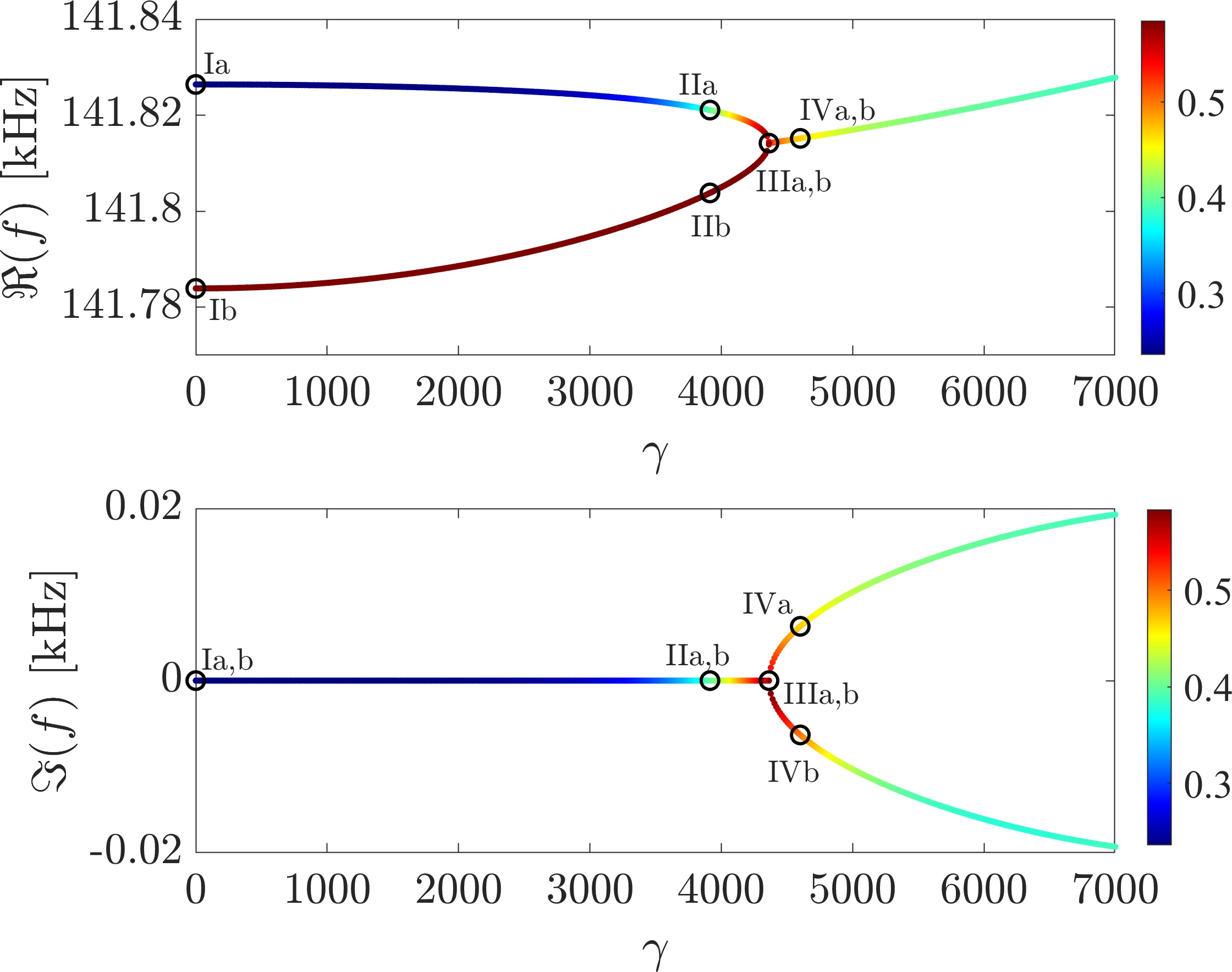}\label{Figsurf2b}}
	\subfigure[]{\includegraphics[width=1\textwidth]{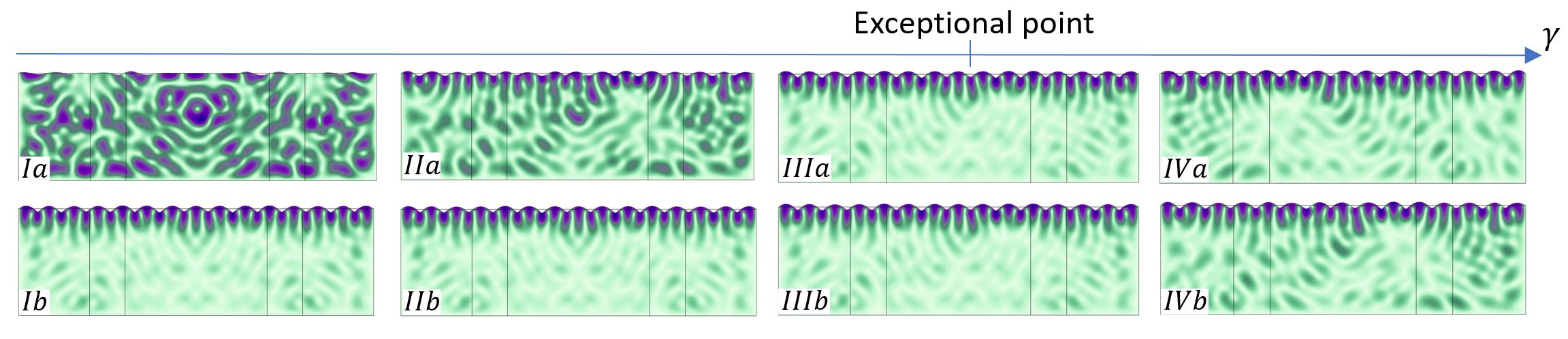}\label{Figsurf2c}}
	\caption{Real and imaginary frequency components as a function of $\gamma$ illustrating formation of the EP (a). The frequencies are color-coded according to the polarization $p_s$, indicating the hybridization of the surface mode (red) and the bulk mode (blue) as the EP is formed, also illustrated by the deformed mode shapes in (c).}
	\label{Figsurf2}
\end{figure}

Figure~\ref{Figsurf2b} displays the real and imaginary eigenfrequency components as a function of $\gamma$ in the frequency range of one EP, formed by a surface wave mode and a bulk mode. The modes are differentiated by a polarization factor $p_s$ defined as
\begin{equation}\label{polsurf}
p_s= \dfrac{\displaystyle\int_0^{L_x}\int_{0.75L_y}^{L_y} (u_x^2+u_y^2) dA}{\displaystyle \int_0^{L_x}\int_{0}^{L_y} (u_x^2+u_y^2) dA}.
\end{equation}
Such polarization factor averages the total displacement at a region near the top surface ($y \in [0.75L_y, L_y]$), which is then divided by the total displacement integrated in the entire domain, and is employed as the color of the dots representing the eigenfrequencies in Fig.~\ref{Figsurf2b}. Hence, higher values of $p_s$ (red) signal modes with energy concentrated at the surface, while lower values of $p_s$ (blue) indicate globally spanning modes, as verified in the mode shapes of Fig.~\ref{Figsurf2c}. The modes concentrated at the surface are formed by a standing rayleigh wave, with a small contribution from other bulk modes. For example, mode Ib in Fig.~\ref{Figsurf2c} corresponds to an eigenfrequency of $f=141.784$ kHz. At that frequency, the non-dispersive rayleigh waves are characterized by a wavelength $\lambda=c_r/f = 2.07$ cm approximately. In the domain with $L_x=30$ cm, this mode should contain $Lx/\lambda=14.5$ wavelengths approximately, which can be verified in its deformed shape displayed in Fig.~\ref{Figsurf2c}. As $\gamma$ increases, the surface mode hybridizes with the bulk mode to form the EP, as indicated by the color changes associated with $p_s$, and also visualized in the deformed mode shapes displayed in Fig.~\ref{Figsurf2c}. At the exceptional point, both eigenfrequencies produce a single linearly independent eigenvector, constituted primarily of a surface wave mode.

\begin{figure}[t!]
	\centering
	\subfigure[]{\includegraphics[height=0.245\textwidth]{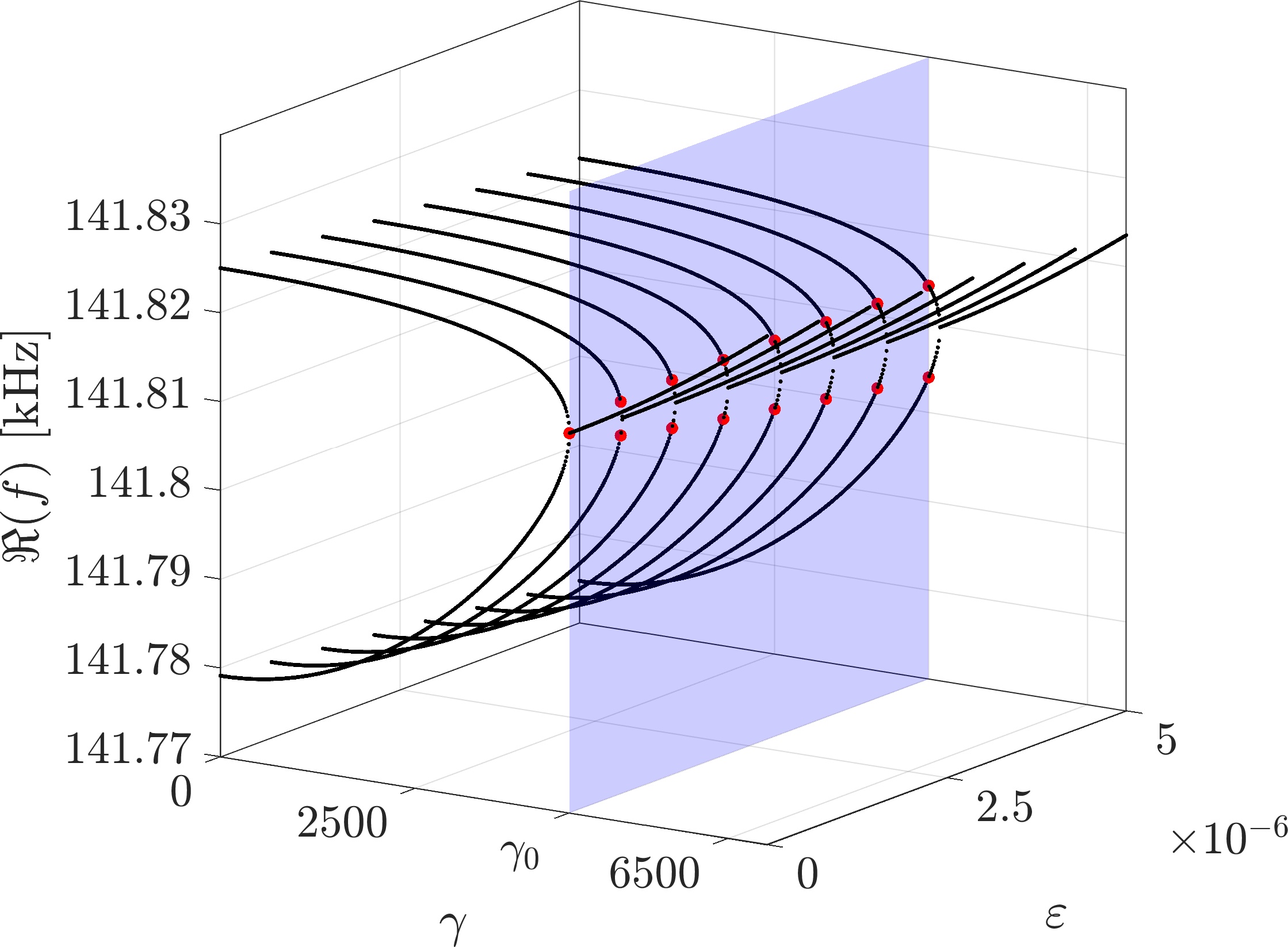}\label{Figsurf3a}}
	\subfigure[]{\includegraphics[height=0.245\textwidth]{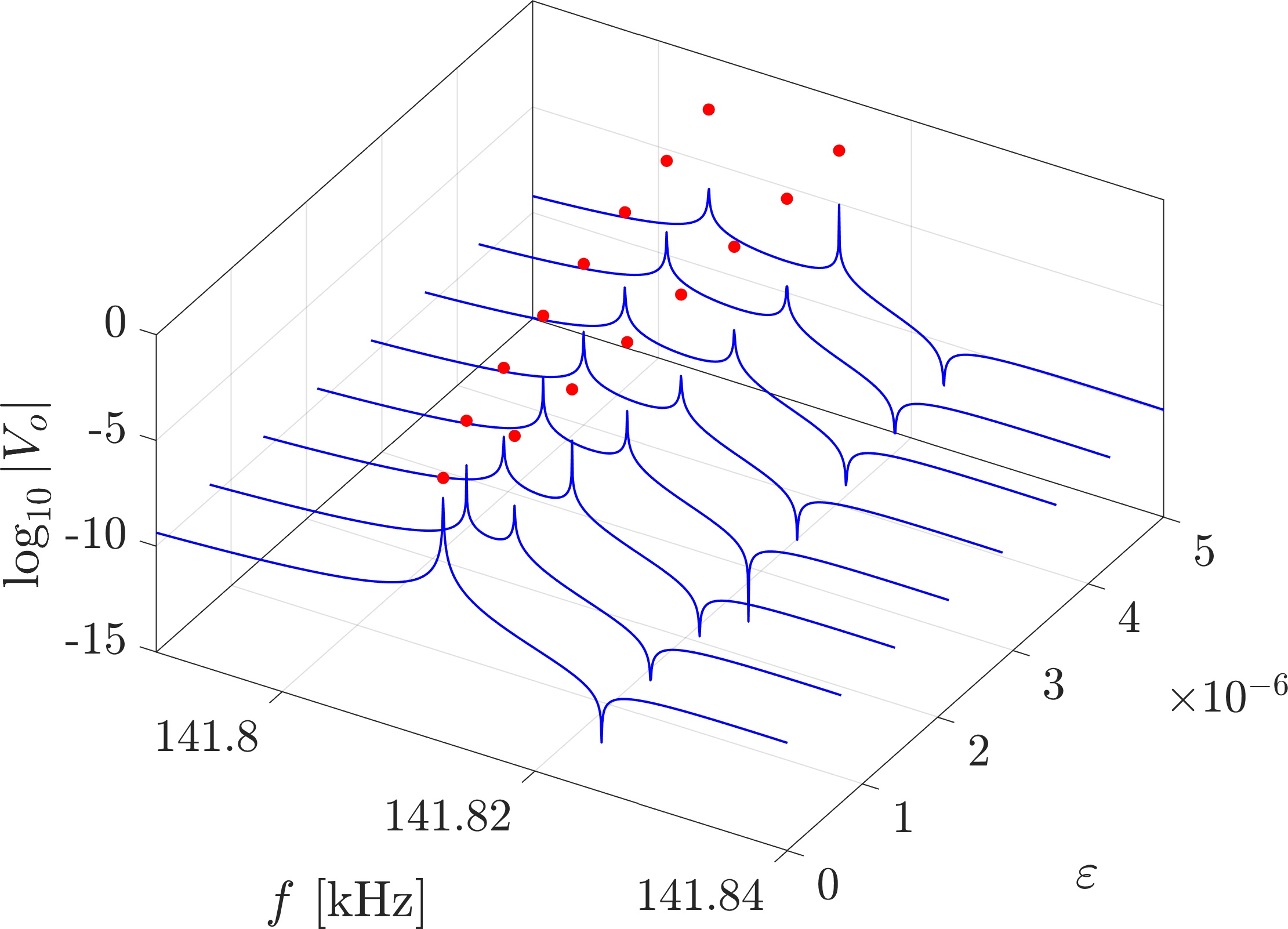}\label{Figsurf3b}}
	\subfigure[]{\includegraphics[height=0.245\textwidth]{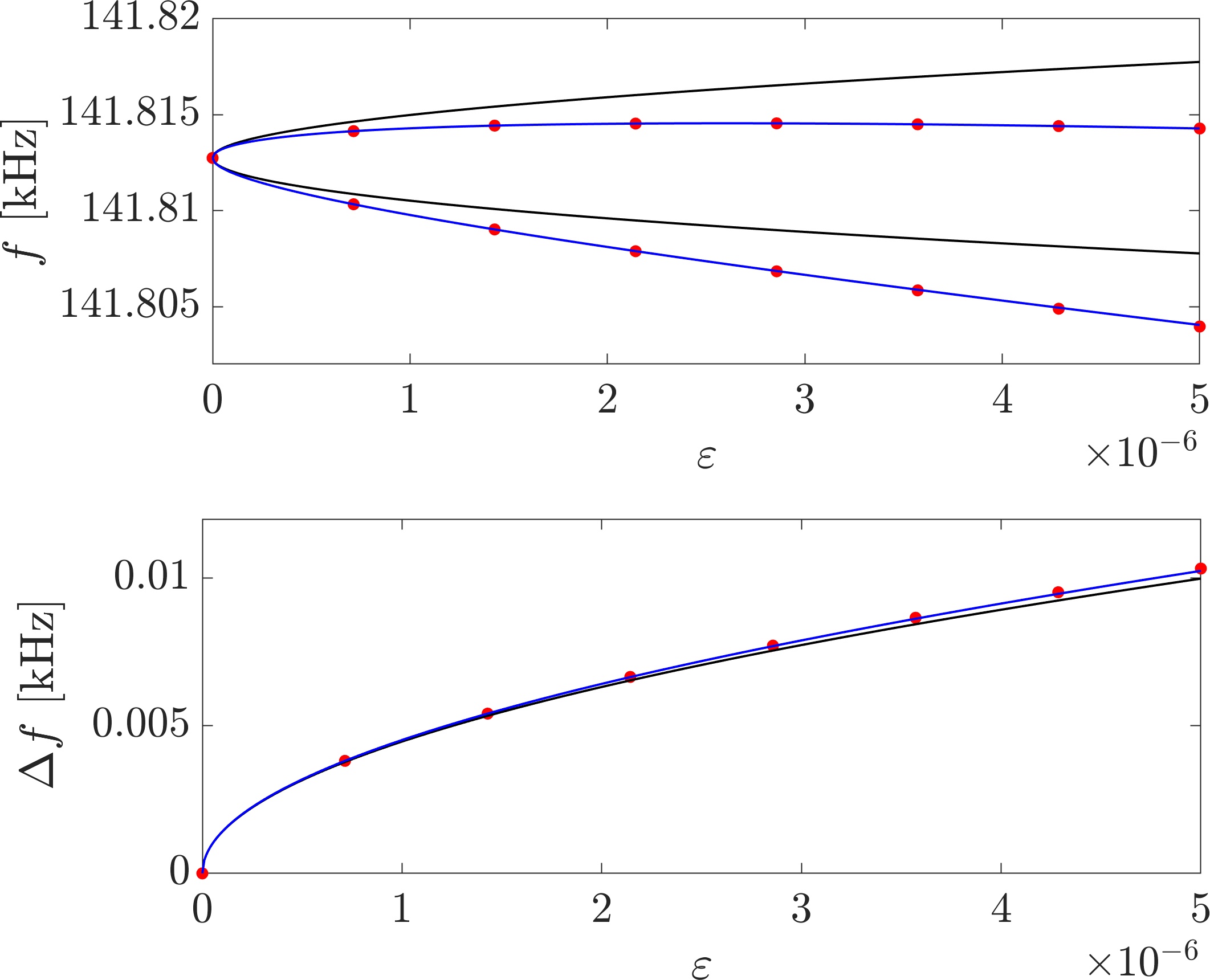}\label{Figsurf3c}}
	\subfigure[]{\includegraphics[height=0.245\textwidth]{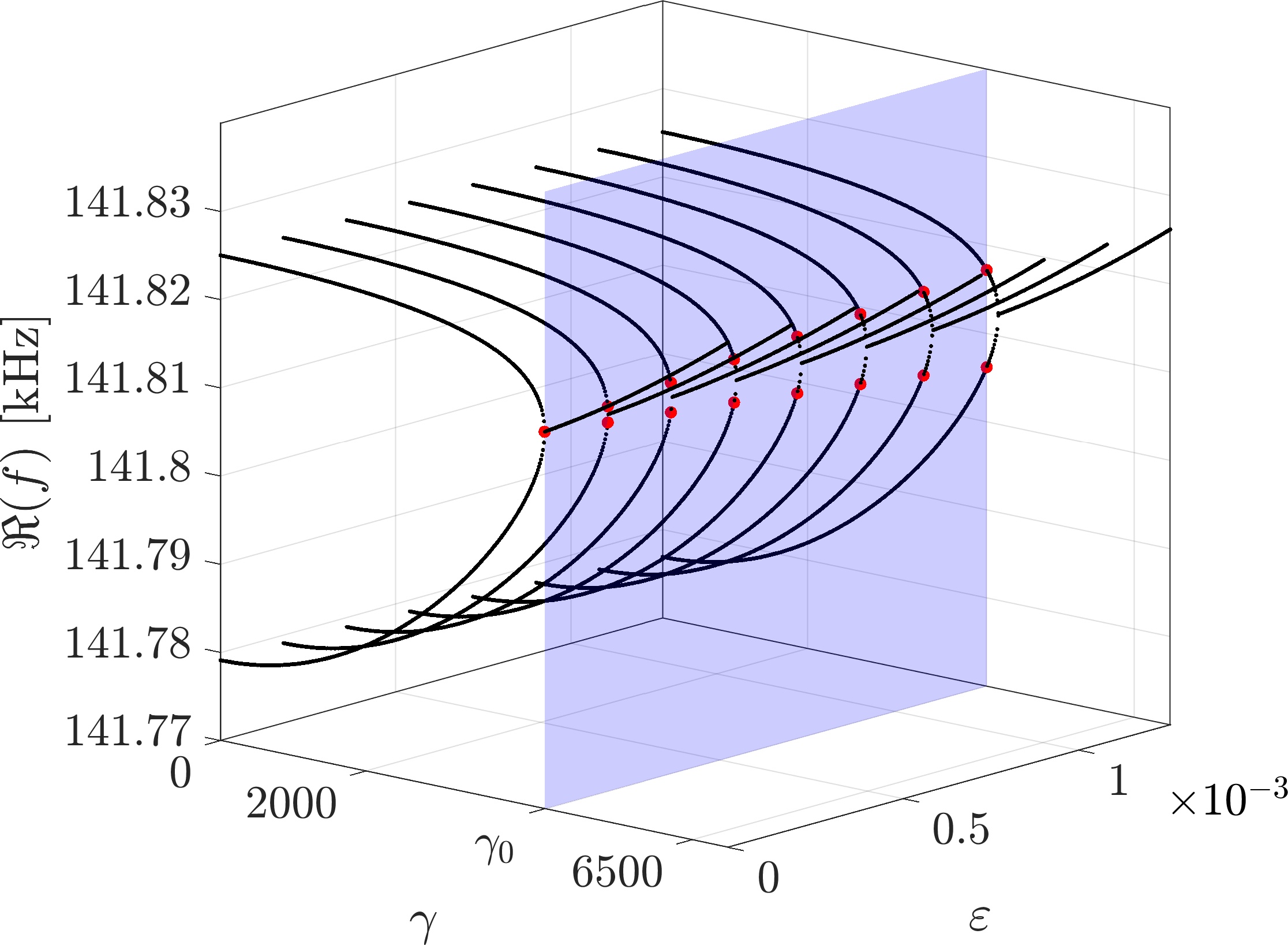}\label{Figsurf3d}}
	\subfigure[]{\includegraphics[height=0.245\textwidth]{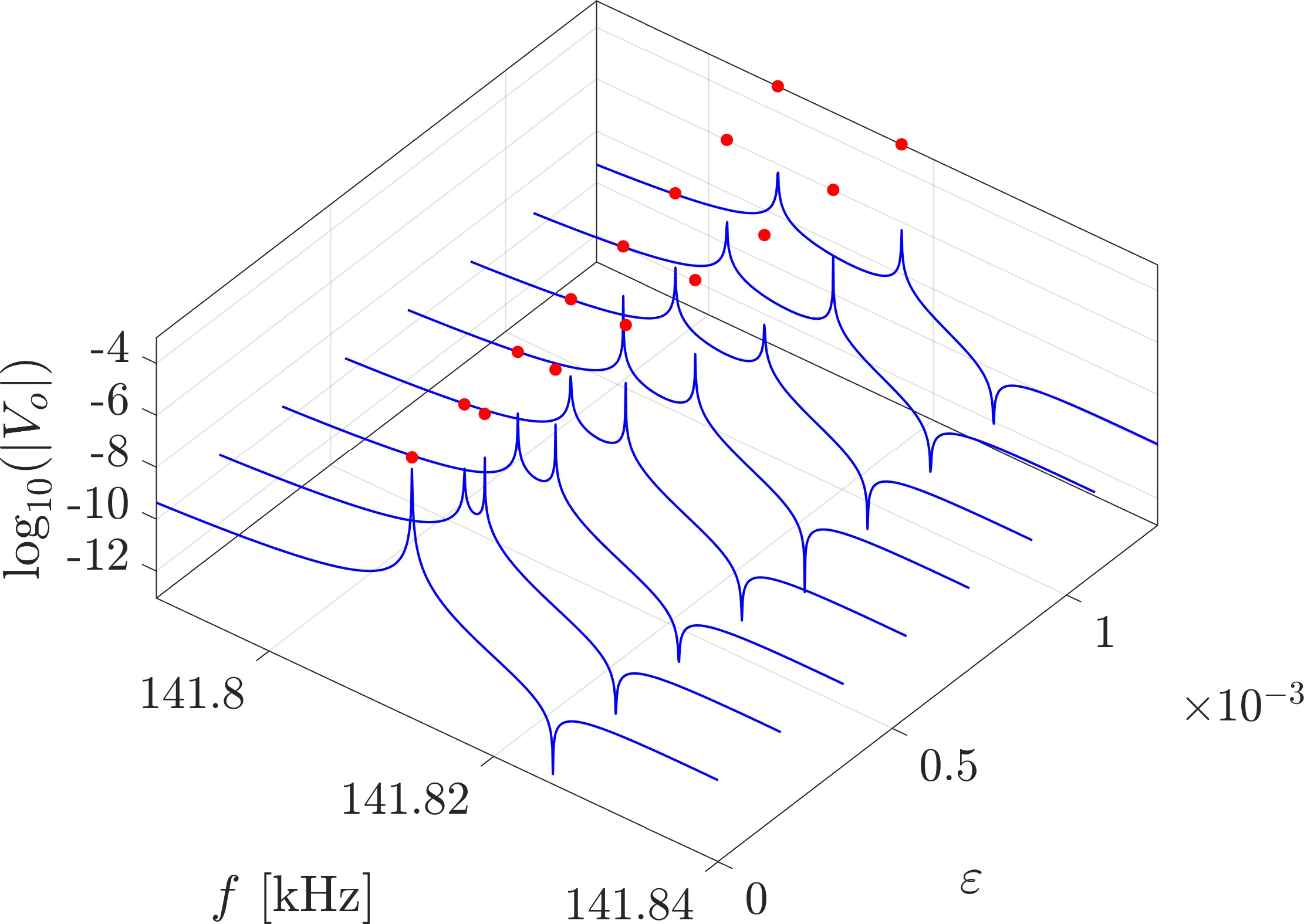}\label{Figsurf3e}}
	\subfigure[]{\includegraphics[height=0.245\textwidth]{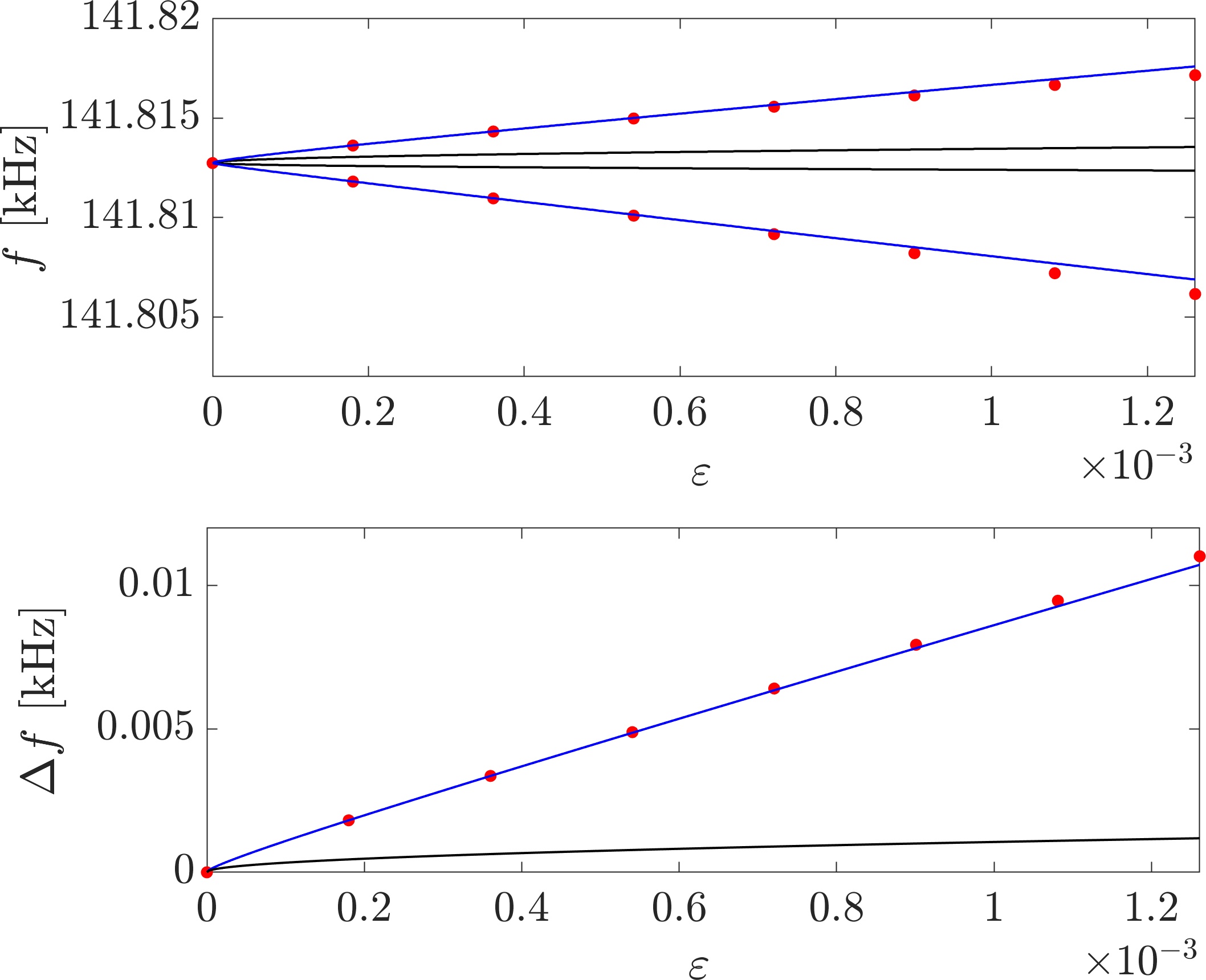}\label{Figsurf3f}}
	\caption{Variation of eigenfrequencies forming the EP with added mass ($\epsilon=M_a/(\rho L_x L_y)$) (a) and crack height ($\epsilon=c_h/L_y$) (d). The blue planes defined by $\gamma=\gamma_0$ highlights the bifurcation of the EP as a function of $\mu_0$ or $c_h$ (red dots). The frequency response plots in (b,e) illustrate the splitting of the resonant peaks measured by the sensor, which match the splitting of the resonance frequencies. The variation of the frequencies and frequency splitting $\Delta f$ with $\epsilon$ are displayed in the top and bottom panels of (c,f), compared with polynomial fits (solid curves). }
	\label{Figsurf3}
\end{figure}

Finally, we investigate the sensitivity of the EP to a point mass inclusions and to a small crack defect. Fig.~\ref{Figsurf3a} displays the variation of the EP with the added mass $\epsilon=M_a/(\rho L_x L_y)$, while Fig.~\ref{Figsurf3d} displays the variation with the crack height $\epsilon=c_h/L_y$, for a fixed crack width of $c_w= 32.43 \mu$m. The blue planes for $\gamma=\gamma_0$ highlights the bifurcation of the EP as a function of the perturbation $\epsilon$ (red dots). Similar to previous results, this eigenfrequency splitting may be detected by monitoring the resonant peaks split (Figs.~\ref{Figsurf3}(b,d)). The corresponding frequency splits $\Delta f$ for the added mass and the crack are displayed in the top and bottom panels of Figs.~\ref{Figsurf3}(c,f), along with solid curves representing the polynomial fits previously described. Compared with previous results in this paper, we note that considerable $\Delta f$ shows a larger linear dependence on the perturbation, and a smaller square root dependence, especially for the case of the surface crack. A dominant square root dependence would be found for even smaller $\epsilon$, which would generate a $\Delta f$ that would challenge the frequency resolution of sensing systems. These considerations will be the subject of future investigations.

\section{Conclusions}\label{Conclusionsec}
In this work we have investigated elastic media such as 1D waveguides and 2D elastic domains, where PT-symmetric attachments leads to the formation of exceptional points in their vibrational spectra. We have illustrated the sensitivity of the EPs to defects such as point mass inclusions and surface cracks. In particular, the defects produce a bifurcation of the EP into two eigenfrequency branches which can be measured through the splitting of resonant peaks in the frequency response. Several opportunities are identified for future studies, such as the design of higher order EPs~\cite{hodaei2017enhanced,xiao2019enhanced} which provide even higher sensitivity, investigations of alternative structures and PT-symmetric gain/loss strategies, influence of noise and non-linearities and experimental demonstrations. Also, this work suggests the monitoring of resonant peaks associated with global vibration modes as a detection strategy, but other methods based on wave scattering ~\cite{xiao2019enhanced} might be of significance for future work on surface waves. 

\section*{Acknowledgments}
The authors gratefully acknowledge the support from the National Science Foundation (NSF) through the EFRI 1741685 grant and from the Army Research Office through grant W911NF-18-1-0036.

\bibliographystyle{unsrt}
\bibliography{References}

\end{document}